   \title{UII Communication Architecture}
   \author{Kari Visala}

\documentclass[12pt,a4paper,titlepage,oneside]{report}

   \usepackage{graphicx}
   \usepackage[left=3.45cm,top=3.5cm,right=2.2cm,bottom=3.5cm]{geometry}     

   \usepackage{epsfig}

   \usepackage{rotating}

\graphicspath{{./}{../graphics/}}

\def\addnotation #1: #2#3{#1       {#2 \dotfill \pageref{#3}}\\}
\def\newnot#1{\label{#1}}

\newenvironment{mylisting}
{\begin{list}{}{\setlength{\leftmargin}{1em}}\item\scriptsize\bfseries}
{\end{list}}

\begin{document}


     \pagestyle{empty}


\begin{titlepage}

\noindent TAMPERE UNIVERSITY OF TECHNOLOGY\\
\noindent Department of Information Technology\\

\vspace{6,5cm}

\noindent {\bf KARI VISALA}\\
\noindent {\bf HYBRID COMMUNICATION ARCHITECTURE HCA}\\
\noindent Master's Thesis\\
\noindent January 17, 2006 \\

\vspace{6,5cm}

\begin{verbatim}
        Subject approved by the department council on 9th February 2005

        Supervisors: Prof. Mikko Tiusanen (TUT)
                     Prof. Tommi Mikkonen (TUT)
\end{verbatim}

\end{titlepage}

     \pagenumbering{roman} \setcounter{page}{1}
     \pagestyle{plain}

\newpage

\addcontentsline{toc}{chapter}{Acknowledgements}
\chapter*{Acknowledgements\markboth{Acknowledgements}{Acknowledgements}}

\begin{quotation}

This thesis is a result of long part-time design process that spans over several years. 
The motivation for this work originates from my work for a platform for distributed shared virtual worlds in Mediaclick Oy
and in a sense I have been pondering on this problem since 1998. 
Technical details have changed along the way but the goals have remained the same.
For many years this work has been in the background
while I have been engaged in other activities, but since I think that despite the fast progress on the field there is 
still a lot of potential in these ideas, I decided to write the design of the communication architecture as my master's thesis 
and see if I am going to have the energy to some day fully implement the system. Or maybe I will find somebody else
interested in this system.
I would like to thank my supervisor Prof. Mikko Tiusanen for his insight and advice in writing of the thesis.
I would also like to thank my parents Ulla and Keijo for their constant background support in everything I have done.

\end{quotation}

\vspace{1,0cm}

\noindent Tampere, January 17, 2006. \\
\noindent Kari Visala


\addcontentsline{toc}{chapter}{Contents}

\tableofcontents
   \newpage

\addcontentsline{toc}{chapter}{Abstract}


\noindent TAMPERE UNIVERSITY OF TECHNOLOGY

\noindent Department of Information Technology

\noindent Institute of Software Systems

\noindent VISALA, KARI: Hybrid Communication Architecture HCA

\noindent Master of Science Thesis, 120 pages.

\noindent Examiners: Prof. Mikko Tiusanen and Prof. Tommi Mikkonen

\noindent May 2006

\footnotesize
The beginning of the 21st century has seen many projects on distributed hash tables, both research and commericial.
One of their aims has been to 
replace the first generation of file sharing software with scalable peer-to-peer architectures.
On other fronts, the same techniques are applied, for example, to content delivery networks, 
streaming networks, cooperative caches, distributed file systems, and grid computing architectures 
for scientific use. This trend has emerged because with cooperative peers it is possible to asymptotically 
enhance the use of resouces in sharing of data compared to the basic client-server architecture.

The need for distribution of data is wide and one could argue  
that it is as fundamental a building block as the message passing of the Internet.
As an answer to this need a new scalable architecture is introduced: 
Hybrid Communication Architecture (HCA), 
which provides both data sharing and message passing as communication primitives for applications.
HCA can be regarded as an abstraction layer 
for communication which is further encapsulated by a higher-level middleware. HCA is aimed at general use, and it is not 
designed for any particular application. One key idea is to combine data sharing with streaming since together they 
enable many applications not easily implementable with only one of these features.
For example, a game application could share the game world state between 
clients and modify it by using streaming.
The other distinctive feature of the system is the use of knowledge of the physical network topology in the optimization of the communication. 
With a feasible business model, fault-tolerance, and security features, HCA is aimed eventually for real-life adoption.

This thesis presents the specification of the C++ client interface of HCA and the architecture and protocol of the distributed nodes forming 
the implementation. HCA is also compared to selected similar technologies and key features of the system 
are discussed.

\textbf{Keywords: scalability, middleware, peer-to-peer protocol, distributed hash table, content delivery network, 
distributed file system.}

\normalsize




\newpage
\addcontentsline{toc}{chapter}{Tiivistelm\"{a}}


\noindent TAMPEREEN TEKNILLINEN YLIOPISTO

\noindent Tietotekniikan osasto, ohjelmistotiede

\noindent VISALA, KARI: Hybrid Communication Architecture HCA

\noindent Diplomity\"{o}, 120 s.

\noindent Tarkastajat: prof. Mikko Tiusanen ja prof. Tommi Mikkonen

\noindent Toukokuu 2006

\footnotesize
2000-luvun alusta l\"{a}htien on syntynyt useita hankkeita kuten BitTorrent ja lukuisia hajautettuja hajautustauluja 
kehitt\"{a}vi\"{a} tutkimusprojekteja, joiden yhten\"{a} tarkoituksena on ollut korvata ensimm\"{a}isen sukupolven 
tiedostonjako-ohjelmistot skaalautuvilla vertaisverkko\-arkkitehtuureilla. Muilla alueilla samoja tekniikoita on 
sovellettu mm. sis\"{a}ll\"{o}n\-v\"{a}litysverkkoihin, hajautettuihin v\"{a}limuisteihin ja tiedosto\-j\"{a}rjestelmiin 
sek\"{a} grid-laskentaan tie\-teel\-li\-ses\-s\"{a} k\"{a}yt\"{o}ss\"{a}. 
Suuntauksen taustalla on yhteis\-ty\"{o}\-h\"{o}n perustuvien vertaisverkkojen ominaisuus 
asymptoottisesti parantaa verkon resurssien k\"{a}ytt\"{o}\"{a} tiedon jakamisessa verrattuna perinteiseen 
asiakas-palvelin-arkkitehtuuriin. 

Tarve tiedon jakamiselle on yleinen ohjelmistoissa ja voitaisiin jopa sanoa sen olevan yht\"{a} perustavaa 
laatua oleva ohjelmistojen rakennusosa kuin Internetin viestinv\"{a}litys.
Ratkaisuna t\"{a}h\"{a}n tarpeeseen on t\"{a}ss\"{a} diplomity\"{o}ss\"{a} kehitetty uusi skaalautuva arkkitehtuuri:
Hybrid Communication Architecture (HCA), joka toteuttaa sek\"{a} tiedon jakamisen ett\"{a} viestinv\"{a}lityksen perusoperaatioinaan. 
HCA on suunniteltu kommunikaation abstraktiokerrokseksi, joka edelleen kapseloidaan korkeamman tason v\"{a}liohjelmistolla.
HCA on suunniteltu yleiseen k\"{a}ytt\"{o}\"{o}n ilman painotusta 
mink\"{a}\"{a}n tietyn sovelluksen suuntaan. Yksi keskeinen idea on yhdist\"{a}\"{a} 
tiedon jakaminen ja tilan streaming-tyyppiset p\"{a}ivitykset yhdeksi primitiiviksi. Yhdess\"{a} n\"{a}m\"{a} ominaisuudet mahdollistavat 
monia sovelluksia, jotka olisi hankala toteuttaa vain toisen ominaisuuden avulla.
Esimerkiksi pelisovellus voi jakaa pelimaailman tilan asiakkaille ja muuttaa maailman tilaa virtaavina p\"{a}ivityksin\"{a}. 
HCA my\"{o}s hy\"{o}dynt\"{a}\"{a} tietoa fyysisen verkon topologiasta p\"{a}\"{a}ll\"{a} olevan virtuaaliverkon 
optimoimisessa. HCA t\"{a}ht\"{a}\"{a} alusta alkaen todelliseen k\"{a}ytt\"{o}\"{o}n ja siit\"{a} syyst\"{a} suunnittelussa on huomioitu 
mahdollinen liikemalli sek\"{a} vikasietoisuus- ja turvallisuusn\"{a}k\"{o}kohdat.

T\"{a}ss\"{a} diplomity\"{o}ss\"{a} m\"{a}\"{a}ritell\"{a}\"{a}n HCA:n C++ asiakasrajapinta 
sek\"{a} j\"{a}rjestelm\"{a}n muodostavien hajautettujen palvelinten 
virtuaaliverkon arkkitehtuuri ja protokolla. Ty\"{o}ss\"{a} verrataan HCA:ta my\"{o}s joukkoon 
samankaltaisia j\"{a}rjestelmi\"{a}, ja j\"{a}rjestelm\"{a}n t\"{a}rkeimpi\"{a} ominaisuuksia analysoidaan tarkemmin.

\textbf{Avainsanat: laajennettavuus, v\"{a}liohjelmisto, vertaisverkkoprotokolla, sis\"{a}ll\"{o}ntoi\-mi\-tus\-verk\-ko, hajautettu 
tiedostoj\"{a}rjestelm\"{a}, hajautettu hajautustaulu.}

\normalsize

   
\newpage
\addcontentsline{toc}{chapter}{List of Terms and Abbreviations}

\chapter*{List of Terms and Abbreviations}

\markboth{List of Notation \hfill}{List of Notation \hfill}

\addnotation ACID: {Atomicity, Consistency, Isolation and Durability}{ACID}
\addnotation API: {Aplication programming interface}{API}
\addnotation CORBA: {Common object request broker architecture}{CORBA}
\addnotation DDOS: {Distributed denial of service (attack)}{DDOS}
\addnotation DHT: {Distributed hash table}{DHT}
\addnotation DLL: {Dynamic link library}{DLL}
\addnotation DSO: {Distributed shared object}{DSO}
\addnotation GLOBE: {Global Object Based Environment}{GLOBE}
\addnotation GLS: {Globe Location Service}{GLS}
\addnotation HCA: {Hybrid Communication Architecture}{HCA}
\addnotation HTTP: {HyperText Transfer Protocol}{HTTP}
\addnotation IDL: {Interface definition language}{IDL}
\addnotation IPR: {Intellectual property rights}{IPR}
\addnotation MTP: {Multicast Transport Protocol}{MTP}
\addnotation OID: {Object identifier}{OID}
\addnotation OSI: {Open systems interconnect}{OSI}
\addnotation QoS: {Quality of Service}{QoS}
\addnotation TCP: {Transmission control protocol}{TCP}
\addnotation TTL: {Time To Live}{TTL}
\addnotation UDP: {User datagram protocol}{UDP}
 \clearpage



     \pagestyle{headings}

   \pagenumbering{arabic}

\chapter{INTRODUCTION}       
\label{introduction}

\section{A Brief History of the Project}

This master's thesis is a product of a design process that started already in 1998. The initial motivation can be
traced to the need for an application framework for scalable multi-user shared virtual environments when the author
was working in Mediaclick company as a software architect. From the beginning, the vision was that 
the future of distributed software lies in distributed operating systems and general purpose middlewares, which are
software layers between the operating system and the applications. They would 
transform the Internet into a collection of modular software components working together. 
The author concluded then
that the numerous virtual reality solutions with their proprietary protocols would vanish quickly when some general domain,
higher abstraction level
distribution platform becomes popular much like the individual graphical user interfaces of MS-DOS programs vanished
after the introduction of multitasking and graphical windowing system provided by the operating system. The idea, in a
nutshell, was that \emph{shared distributed virtual worlds} could become a uniform user interface to the whole Internet, and users
could work together in artificial environments where distributed software components create their own user interfaces. 

\newnot{UDP}
\newnot{TCP}
No sudden paradigm shift has happened, at least not in the general desktop software. 
For example, lots of software still use  
relatively low-level proprietary protocols on top of TCP \cite{TCP} and UDP \cite{UDP}. 
Typical programs are monoliths when viewed externally, 
the distinction between programs and
files has not been blurred into objects, and local and global resources are often used differently. 

\newnot{IPR}
There have been and still are some difficult
obstacles hindering the progress, like the huge amount of legacy code and the need for backwards compatibility and 
unwillingness of large software intellectual property rights (IPR) owners to open and modularize their main asset.
Also, the general hostility of the environment in open networks and the still rather modest Internet bandwidth and reliability 
when compared to access to local devices.
On the other hand, the proprietary virtual reality software platforms for example, have mostly remained isolated
toys often without proper support for scalability, security, location of resources, and other orthogonal 
high-level features that should, logically, be implemented by a general purpose platform.

After Mediaclick company was placed in liquidation in year 2002, 
the idea of a development of a general purpose distribution platform has been a hobby project of the author. This
has been very much a part-time activity and mostly consisted of just following the progress of the field. 
Basically, all details have 
changed during these years from the original system developed in Mediaclick but when the ideas started to show some 
convergence the author decided to write the first part of them down as this master's thesis. 

\section{Overview}

Despite the many problems facing the development of distributed software for the Internet, there are still many attractive
reasons for developing middleware or operating system for distributed components and having the more abstract
perspective of network of computers as a single large computer containing distributed resources:

\begin{itemize}
\item
The number of potential applications that could benefit from easier communication with
other components in the network and applications that are distributed by their very nature
will steadily increase. The reasons for this include
the growing bandwidth and increasing reliability of Internet as network technologies mature, 
increasing amount of content in the Internet, growing number of
users and computers connected to the Internet, and better awareness of the possibilities of distribution in general.  
\item
Raising the average abstraction level of building blocks used by programmers trades hardware resource utilization to
programmer's time. As hardware gets cheaper and faster, the relative price of the programmer's work rises. Higher level
of abstraction could also grow the number of people able to program.
\item
Smaller, modular distributed components that rely on stronger shared assumptions can have surprising cumulative
benefits when combined in new ways. There could be some critical mass of these components that, when crossed,
the efficiency of writing software on this kind of platform could suddenly pass the benefits of building on legacy code.
\end{itemize}

\newnot{CORBA}
It seems that the benefits of distribution middlewares are generally accepted. Numerous platforms have been 
developed for distributed
software in the broad sense including the biggest companies in software business. 
These include Object Management Group's CORBA
\cite{CORBA}, Sun Microsystems' Java 2 platform Enterprise Edition \cite{JAVA}, 
and Microsoft's DCOM \cite{DCOM} and .NET \cite{DOTNET} frameworks. 

Many of these systems could
be counted in one of two categories based on their communication primitive used: Some provide a distributed shared
memory with certain consistency properties for their applications, but most of the systems are based on 
object-oriented entities that communicate by sending messages. 

Distributed objects are usually implemented
as having a message "mailbox" and invocation of operations of the object are mapped to messages sent to the object.
This way of mapping objects to networked computers has the benefit that logical encapsulation of the implementation
shares the physical boundary of a computer system. This provides security in a natural way, because an object implementation
can only be accessed through its interface. Also, it is usually possible to point out an organization that has the responsibility
of administering each object in the system.
Because there can be many clients for a single object we could characterize this type of communication as many-to-one.

A typical implementation for shared memory is just the opposite: Usually the writing to a certain address space is 
restricted to one entity at a time but many clients can read the memory simultaneously. Therefore, it could be called
one-to-many. It also seems that sharing of data is most naturally applied to the distribution of finite datatypes that do
not hide their implementation structure and can be stored in finite space. For example, the type of list of integers
can be implemented as an algebraic datatype having two constructors \verb1Empty : () -> List1 and 
\verb1Cons : Integer -> List -> List1.
Every element of this type is finite and can be stored in memory and used by arbitrary functions in clients.

These two basic communication patterns can obviously be implemented on top of each other: In a message-based system it is 
possible to implement an object providing a shared state for other objects and it is possible to use
shared memory for passing messages between objects. However, the platforms are typically biased in their design for 
one of these paradigms and for application designs it is much easier to follow the native model of the underlying
layer consistently. The middleware layer also hides the underlying network technology and implementation
details of the communication primitives from the applications, which means that even if one of the paradigms of 
communication is implemented on top of the other, the implementation will not map efficiently to the underlying resources.
The shared memory primitive and the messaging primitive implementations can also require a substantial amount of
work when done properly.
Also, conceptually they both reside on the same abstraction level, which means that their full potential is not harnessed 
if one of them is implemented on application level 
and the other on platform level. Therefore, we start from the basic assumption that if both styles of communication are
commonly used, then both communication primitives 
should be integrated to the implementation of the distribution middleware. 

It can be argued that many-to-many communication primitive is not needed at the same abstraction level as many-to-one and
one-to-many, because, it can be implemented as
a combination of many-to-one and one-to-many with a controller component in the middle, assuming that the controller component can
handle the bandwidth generated by multiple producers. 
On the other hand, if the bandwidth
cannot be handled by a single component, the receivers of the channel should have some means of filtering the data and it is
unclear if there is any design criteria for selecting the method of filtering.
However, it should be noted that some systems 
might have an approach totally different from both of these communication primitives and there does not necessarily exist
a simple one-to-one mapping from application programming concepts to the underlying implementation concepts.

\subsection{Categorical Datatypes}

Two programming paradigms have remained popular in many modern programming languages: 

\begin{itemize}
\item
Object-oriented style (see \cite{OLIOKIRJA} for general introduction), where
programs are composed of encapsulated objects having a fixed interface and interaction with the object is performed through
that interface. This structuring of a program is particularly suitable for entities that can have multiple implementations
but fixed interface or fixed ways to use the entity. Especially infinite datatypes like infinite lists are naturally modeled
by objects. For example, infinite list of integers can be modelled as an interface containing two destructors
\verb1First : List -> Integer1 and \verb1Tail : List -> List1.
\item
Programming style based on algebraic datatypes is used heavily in functional programming languages like Haskell \cite{HASKELL} and ML \cite{ML}.
These datatypes are defined by a fixed set of constructors for building the objects of the type and then writing
functions that handle these datatypes. These datatype definitions are the opposite of object interface
definitions, because with algebraic datatypes, the implementation of the objects of the type are fixed, but it is
easy to write new functions that operate on these types.
\end{itemize}

Categorical datatypes based on the \emph{dual} concepts of \emph{final coalgebras} and \emph{initial algebras} of category theory were
introduced in 1987 by Hagino in the programming language Charity \cite{hagino87}. It is known that initial algebras
can be used as a formalism for describing algebraic datatypes, and in 1995 Reichel showed in \cite{rei95} that
object semantics can be mapped to final coalgebras. These ideas have then been developed further by many researchers
and, for example, in \cite{poll98subtyping} Poll showed how to add subtyping and a simple form of inheritance to 
categorical datatypes and concluded that coalgebraic types can provide a useful calculus for encoding of objects.
Generally, it seems that category theory is a good formalism for describing datatypes and reveals a surprising 
symmetry between the two popular programming paradigms above.

\subsection{Hybrid Communication Architecture}

Because objects and algebraic datatypes are natural structures for different types of entities and because of their
symmetry we will subscribe to
the view that a programming language would benefit from a good support for both of them. It is also interesting that
while object invocations map naturally to messages in a distributed implementation, objects of algebraic datatypes 
are efficiently communicated by sharing their actual data with all interested receivers. This is because the representation of
the objects of algebraic datatypes is known and can be stored in finite space.

The above can be interpreted that network should also provide distributed memory resources in addition
to message passing to efficiently and naturally facilitate the use of these dual programming styles.
These two basic functions of network should be viewed equally important and residing at the same abstraction
level. Building higher-level abstractions on top of these primitives has the potential to lead to a balanced design and
efficient harnessing of physical resources. The emphasis on both distributed state and messages is 
the reason for the name \emph{Hybrid Communication Architecture} (HCA).
\newnot{HCA}

\newnot{OSI}
In this thesis we will introduce HCA, which is the lowest-level layer of distribution middleware/distributed
operating system. We will basically start from scratch from the application layer
of the OSI \cite{OSI} reference model. The implementation of HCA is probably first built on top of TCP and UDP protocols but HCA should
also be implementable on other network technologies. 

HCA is aimed at abstracting away
the network technologies and heterogeneous operating systems currently populating the Internet and functioning as
an abstraction layer for communication providing efficient, scalable,
location transparent and fault-tolerant primitives supporting migrating clients
for message passing and sharing of data. The HCA interface to higher layers should be as simple as 
possible without significantly making worse the utilization of underlying resources. The goal for the next layer on top
of HCA is to raise the abstraction level for writing applications.

It was decided to combine \emph{push-style streaming} to the data sharing primitive because they are easy to implement 
together and enable an efficient implementation of several applications that would be cumbersome to implement with only
streaming or data sharing. This means that after data has been published to the network, it can be modified 
afterwards with smaller modification packets and interested entities are signalled automatically about these
updates.

\newnot{DHT}
The scalability of the architecture is achieved by relaxing transactional properties of the shared state and
implementing the system as a network of communication nodes that
implement an efficient overlay topology for distributing data by applying multi-cast-like techniques, \emph{distributed
hash tables} (DHT)\cite{CHORD} and extensive caching. Fault-tolerance is achieved by using replication. 
This complexity is completely hidden from the clients of the HCA.

The stateful approach to the communication architecture, where the middleware has an internal state, has also its drawbacks. Maintaining a consistent state
over concurrent communication nodes makes the implementation and protocol semantics more complex and error-prone.
For example, CORBA has chosen the stateless approach
as the basis of its communication architecture partly for these reasons. 
Also, storing cache replicas of data to the edge of network raises problems in access control and security of the data.
However, to fully exploit network resources, a stateful model and extensive caching is needed. Without caching the clients that wish to share
data with other clients need to send the data multiple times from end to end to each client interested in the data.

\section{Possible Applications}

Although HCA is designed for general purpose use and to serve as a system abstraction layer for communication in
a part of larger framework for building distributed software, it has also value as an individual component. We will list
some interesting and natural applications of the system here:

\newnot{HTTP}
\begin{itemize}
\item
The scalable sharing of data could be used, for example, in multi-user games to distribute the state changes of the
game world efficiently to multiple receivers. The same technique can be applied in virtual environments and simulations.
\item
HCA also has all the basic features of a distributed file system and it could be used for storage and fast
retrieval of large amounts of data. HCA could even be used to replace HTTP to implement a kind of "super-WWW"
with push semantics and scalable data distribution where publishers of information do not need upload bandwidth
proportional to the number of readers of the information.
\item
Scalable streaming of data could be used to implement a symmetric video chat application where
streams from any access points could suddenly be distributed to millions of interested receivers.
\end{itemize}

\section{Contents of this Thesis}  

This thesis presents the design of basic architecture, algorithms, and protocol of HCA. The general 
principles guiding the design are also discussed and the results compared against these criteria.

In Chapter \ref{ch:problemDef}, we will further examine the problem of scalable distribution of data.
The most important aspects of data distribution are analyzed and criteria for evaluation of the work are set. Also,
a survey of existing technologies in the field is made and some representative systems are introduced briefly.
In Chapter \ref{ch:architecture}, the Hybrid Communication Architecture is presented at general level as a solution
to the problem of combining a scalable data sharing and a message-based primitive into a system abstraction layer for communication.
In Chapter \ref{ch:interfaces}, the C++ programming interface exposed to HCA client programs is presented.
In Chapter \ref{ch:domaintree}, the overall structure of the HCA network is presented.
In Chapter \ref{ch:protocols}, the protocol used by HCA communication nodes is specified.
In Chapter \ref{ch:discussion}, the protocol is analyzed and the results are discussed and matched to the criteria first set in Chapter \ref{ch:problemDef} 
and the designed system is compared to selected other technologies.
In the final Chapter \ref{ch:conclusion}, a conclusion of the work and future possibilities and 
directions for the work are discussed.


   \chapter{SCALABLE DISTRIBUTION OF DATA}
\label{ch:problemDef}

\begin{figure} [ht]
\centering
\centerline{\epsfxsize 0.45\linewidth \epsfbox{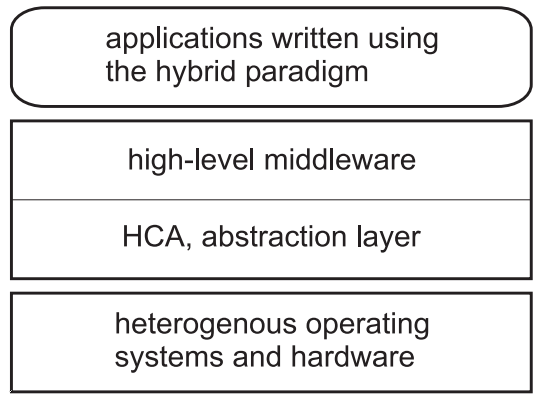}}
\caption{A general architecture of the system}
\label{archdiag}
\end{figure}

The goal of this thesis is to present an architecture, 
an interface, algorithms, and a protocol for hardware and operating system abstraction layer for communication that can be used
to build a potentially global high-level distribution middleware supporting both object-oriented and functional programming
styles. The general level layered architecture of the system is shown in Figure \ref{archdiag}. The specified system should be a stand-alone component and 
designed for commercial use from
the beginning. There can be many implementations of the architecture for different network and computer technologies,
operating systems and programming languages. It should be possible to combine components from multiple implementations to form a single system 
because we specify the interfaces and protocols between components of the implementation.

The system should be as simple as possible without significantly sacrificing the efficient utilization of underlying resources when
used for message passing and sharing of data. The data sharing primitive of the system should also support push-style
updates to the shared data so that updates are distributed with small latency to the receivers. Both communication
primitives needed implement best-effort semantics on top of possibly unreliable networks so that connections may 
be broken, for example, in case of global network partitionings, but clients are always notified of these errors, and 
system guarantees that no communicated information is lost without notification. Also, all delivered information is
received unchanged and system should recover from many types of failures transparently.

\section{Non-functional aspects of a distribution platform}

In the following sections, different desirable general features of distributed system are reviewed briefly.
These features are aspects of the system as whole and are in many cases independent of 
the actual interface of the system. We have used the aspects presented here as general design goals of HCA and
the results are evaluated in Chapter \ref{ch:discussion} against these criteria. We will concentrate on
scalability, location transparency, locality, security, fault-tolerance, and feasibility of global implementation of 
the system in this thesis, but other criteria presented here are also considered in the design. Compatibility with
other systems has not been among the goals of this system and, basically, we have designed it from scratch to leave
as much flexibility as possible to the implementation of the other features.

\subsubsection{Scalability}

The most important goal in designing HCA has been the scalability of the system. 
In particular, the scalability of the architecture, cost, and efficiency of the implementation as a function
of the number of clients assuming each client has limited bandwidth available for both sending and receiving data. Clients
can be understood here as computers or programs using a HCA implementation.
This means that as the number of clients of the system increases, the following statements hold true:

\begin{itemize}
\item 
The same architecture and solutions can be used independently of the number of clients. Especially, the implementation
should be possible to be composed of standard components with constant properties and the scalability should be
achieved by increasing the number of these components. This applies also to the setup, configuration, and maintenance
needed by the system.
\item
We should be able to achieve nearly $\Theta(n*log(n))$ asymptotic total cost of building and using the system as a function of
the number of clients $n$.
This is sufficiently close to a constant cost per client. 
\item
Efficiency of the system seen from a single client perspective is not affected by the number of other
clients. It should be possible to achieve $O(log(n))$ latency and
$\Theta(1)$ bandwidth as a function of the number of clients $n$.
Also, the other properties like robustness visible to clients should in general be independent of the number
of clients.
\end{itemize}

HCA is aimed for global use and it should be scalable to a situation where $10^{10}$ users use the system
simultaneously and there are hundreds of programs running for each user using HCA. With this kind of scale it
is probable that the underlying network bandwidth will be the first bottleneck of the system, but when
the underlying physical network supports as massive a use as this, HCA should scale along.

\subsubsection{Distribution transparencies}

The ability of a distribution systems to hide many complexities involved in the realization of the system from
the applications can be described with many so-called \emph{distribution transparencies}. We will list the most commonly
used distribution transparencies here:

\begin{itemize}
\item
\emph{Access transparency} means that the actual mechanism of invocation and data representation are hidden from the application.
For example, local and global resources are accessed in the same way.
\item
\emph{Location transparency} masks the location information and enables contacting a target without knowing its location.
\item
\emph{Migration transparency} is used in this thesis to mean that a target can change its location while communicating
with an application without the application noticing the migration.
\item
\emph{Failure transparency} masks failures from the application. The implementation can, for example, initiate a protocol for
recovering from a network failure but application is unaware of the failure at all.
\item
\emph{Replication transparency} hides the implementation technique of replicating many copies of an entity in order to
provide reliability and availability of the entity.
\end{itemize}

The more transparencies the system provides, the more abstract is the interface to the system. Usually, the drawbacks
for adding transparencies are a more complex implementation and decreased performance. The more transparencies a system
provides, the closer is the logic that can be used to implement a distributed system to that of a non-distributed system.

\subsubsection{Simplicity}

An important goal of design of HCA is simplicity in both the interface and implementation of the system. The interface
of the system should provide the absolute minimum of features required for an efficient implementation to achieve good
modularity. It is the goal of the next layer on top of HCA to raise the abstraction level of the system. When HCA is modular,
it is easier to implement and implementation can concentrate on scalability and efficiency.

The internal architecture of the HCA should be as simple as possible because distributed systems with partial failures and
many parts working together are already very complex by their nature. The simpler the implementation is, the lower is the
probability of errors, unexpected functionality, problems caused by hidden assumptions, and potential security holes.

\subsubsection{Performance}

Because HCA is a relatively low-level layer, it should have a good constant factor efficiency in addition to scalability, 
because it affects all systems built on top of it. 

\subsubsection{Fault-tolerance}

An unreliable underlying distributed hardware where communication between nodes can break and computers can crash
independently while other parts continue working is the typical assumption for distributed systems. These failures can
occur simultaneously and in any kind of combinations. It is, for example, possible for the network to partition temporarily into
separate pieces which continue working independently but are unable to interact with the other pieces.

HCA is expected to work continuously on top of this kind of unreliable hardware. 
The system must be able to recover to a consistent state with high
probability from any kind of combination of failures that is likely to happen. 

HCA has also the goal of failure transparency which means
that the interface of the system should try to mask failures from the clients while the implementation can use sophisticated methods
for recovering from these failures. Underlying failures can affect the performance of
the system seen by clients, but the basic functionality of the interface should be unaffected by them.

The robustness against failures is usually achieved by some sort of redundancy in the implementation. HCA uses replication
of resources as a common solution.

\subsubsection{Locality}

Using local resources should always require only local communication,
local caches should be preferred to distant ones and when using distant resources messages should travel as directly as possible in the underlying
network and geographical topology. 

Communication bandwidth and the speed of memory accesses typically decrease quickly as we move away from the processor: The internal cache of
the processor is faster than an external cache that is faster than memory that can be orders of magnitude faster than accessing a hard disk that
can be faster than local network which can be faster than a company intranet which can be faster than an internet service provider internal
network that can be faster than global Internet access. Latency of communication has also the lower bound dictated by the speed of light and, in practise, correlates
strongly with the number of devices in the communication pathway. Therefore, exploiting locality can have a significant impact on efficiency.
Because geographical location and underlying network topology often reflect the organizational structure, locality in communication can also 
improve the security of the system as many local messages never leave the company intranet. Also, local communications usually have
better reliability than long distance communications, because there are less devices that can fail in the pathway and less interference
from other uses of the same components. 

\subsubsection{Business Model}

Scalable protocols for data sharing are often based on using the whole network for the distribution of popular
data and interworking of participants in the communication. Complex protocols can produce a good asymptotic 
efficiency but at the same time require global coordination in building of the network and use
of resources in it. This is not a problem in small-scale systems, which are managed by a single organization owning
and building the network. In that kind of setting it is plausible to optimize the network topology globally; 
providing fair sharing of resources is easier because the whole network is owned by the same organization. The problem
arises in large-scale systems that must serve multiple organizations. 

A large-scale architecture must support 
fair sharing of resources and building the whole system in independent parts inside organizations if it is 
to be built at all. For example, if the building of the system needs global coordination between the organizations,
it will affect the adoption of the technology. Also, if investing resources to building the system
does not give proportional benefits to the investor, there is little motivation for organizations to participate in building
the system. On the other hand, a system with a better global asymptotic efficiency could eventually force everyone to use the system 
as an altruistic "tribe" if a critical mass of users is attained because the fair system cannot compete with it.

\subsubsection{Support of Transactions}

In database systems, data is transformed from a consistent state to another by \emph{transactions} \cite{TRANSACTIONS}.
Transactions are the isolated logical operations that change the state of the database. Each transaction
can contain multiple tasks and in a distributed setting this can mean changing the states of many distributed databases. 
The relevant properties of transactions are:

\begin{itemize}
\item
Atomicity: each transaction
is guaranteed to be atomic, which means that either all tasks of the transaction are performed or none. 
\item
Consistency: transaction always leaves the system in a consistent global state. 
\item
Isolation: the intermediate states
of transaction are not seen by other users and that the transaction history of the system is serializable. 
\item
Durability: after
a user has been notified of a successful transaction, the changes to the state of the system are persistent and protected from failures.
\end{itemize}

\newnot{ACID}
Together these are called ACID properties.
Transactional semantics can be used in distributed systems and the state of the system is considered to span over multiple
hosts in the network. Algorithms have been developed for implementing transactions in distributed systems, like the
two-phase commit protocol \cite{TRANSACTIONS}. While these can guarantee the ACID properties and raise the abstraction level of the
system considerably, they are complicated to implement, produce new kind of failures (like deadlocks) to the system, can slow
down the system, and contradict with scalability.

In the design of HCA we shall sacrifice a global consistent state for better efficiency and scalability. We have, for example, limited
the number of simultaneous writers of a given data entity to one. There is also no global consistency between different data entities
in the system which means that they can be distributed in isolation. Atomicity, isolation and durability are easy to implement
after this compromise. If stronger consistency is needed, it can, hopefully, be coordinated on a higher level of the architecture.

\subsubsection{Security}

In distributed systems implemented in public networks, security against malicious use is of paramount importance.
This is especially true for large peer-to-peer type architectures where it is necessary to be able
to make changes to the server network flexibly. In a way, powerful features that can affect the
network of nodes globally increase the vulnerability of the system. Because robustness against
attacks is considered more important than the ease of configuration in the design of HCA, some 
compromises have been made. The security requirement also contradicts efficiency. For example, encryption
and authentication of messages to protect communication between nodes consumes processor time. The architecture is, however, designed so that
if an implementation is installed in an already secure network (for example, private physical network), 
it is possible to omit the encrypting of messages.

The guiding principle behind the design of security of HCA has again been scalability. This means that 
even successful attacks should have only local effect on the network. Also, after incursion
has been detected and the security hole removed, the recovery of the system should be possible with a reasonable
effort.

Different types of attack against a network of communicating nodes can be categorized as follows:
\begin{itemize}
\item changing the topology of the network by adding or removing nodes,
\item using a compromised node to break the interworking between nodes,
\item attacking the system by using it in a basically legitimate way (say, DDOS attacks, as explained below), \newnot{DDOS}
\item by interfering in communication between nodes by adding, removing, changing, or eavesdropping of messages.
\end{itemize}

The best protection against compromised but trusted nodes or adding and removal of nodes is to have
a robust and decentralized protocol without any nodes implementing any kind of central control that could function as 
a single point of failure.

It is possible to try to harm the system by using it through its interface.
Distributed denial of service (DDOS) attack refers to a form of attack where multiple computers
all over the network flood one target computer by sending as many messages as they can. This prevents
the legitimate users of the target computer from using its services. The attacking computers are
first compromised, for example, with some kind of virus program. The actual attacker then controls the
attack from some point that controls all compromised computers that do the flooding. For DDOS
to succeed there must be a number of vulnerable machines in the net in the first hand. In the case of HCA,
the target of DDOS attack can be one of the HCA communication nodes or, more probably, some client using HCA
and the attack is just implemented through HCA network. It is difficult to devise protection from
a DDOS attack if there are lots of compromised client machines, because the attack cannot be easily distinguished from normal
usage of the system. 

The protection from interference of communication between nodes has many well-known solutions.
Encryption, error coding, and digital signing of messages
makes it impossible to change, add, or eavesdrop messages between nodes assuming that keys remain secret and encryption method is not broken 
algorithmically. It is still possible to remove
messages from the underlying network, but this manifests itself to the system in the same way as normal
loss of messages in an unreliable network.

The attacker could just try to sabotage the system or use resources of the system unfairly. The latter
possibility should be taken into account in the design of the protocol. The attacker could also target
the attack against a particular user of the system and not the system itself. This could be done, for example,
by stealing the identity of the user and using the identity to read, change, add, or remove information owned by the user,
or masquerading as the user. It should be noted in the design that when the network of nodes is 
built by multiple organizations, these organizations can be hostile against particular users of the system or the organization 
itself could try to benefit from compromising the semantics of the HCA, for example, by reading information protected by access rights.
By selecting trusted organizations to administer the HCA network, a certain level of security can be expected, but already the large
number of nodes makes it likely that some subset of them are compromised.

Additional techniques to add security could be a data backup mechanisms for adding protection against
changing and removal of data and logging and monitoring of usage for accountability.

\subsubsection{Maintainability}

Maintainability is used to refer to the total work needed to set up and maintain a working installation, assumptions about
the organizational structure and environment required by the implementation, and flexibility to a change in environment of the system.
An ideal system to maintain is built on unreliable
hardware like home computers, doesn't need any configuration from users, and different parts of the system are provided by different
organizations that are potentially hostile to each other. This kind of system resembles almost a living organism that continues
working in a constantly changing hostile environment. This has been the goal of many peer-to-peer systems ran on home computers
without any central coordination. 

\subsubsection{Compatibility}

Compatibility to other systems and incremental development are important goals for software systems aimed at commercial use. In the context of 
distributed sharing of data, an easy integration to web technology would be very beneficial and speed up the introduction
of the system. Compatibility has not been a priority in the design of HCA because the aim has been to eventually create a very
high abstraction level middleware for composing distributed software and, therefore, the design has started from scratch to avoid
unnecessary dependencies.

Because HCA will have well defined interfaces, there can be multiple implementations of the components of the architecture.
These different implementations should be possible to combine to form a single HCA overlay network.

\subsubsection{Quality of Service}

\newnot{QoS}
Quality of service (QoS) is a feature of a non-ideal world: Because properties of communication like bandwidth, latency, security, and reliability
are not free and network resources are not abundant, we have to sometimes prioritize the use of available resources or to have some sort of guarantees
from the system that a certain level of quality of service is provided. The first version of the HCA does not provide support for 
guarantees like maximum latency or minimum bandwidth. 
Because HCA implementation will be first targetted on Internet, we would not even have the technology to guarantee these qualities. Also,
the scalable routing algorithm, which depends on cooperation of communication nodes and the need to guarantee fairness to all users, makes
the problem of QoS guarantees even more difficult.

\subsubsection{Anonymity}

Sometimes it is desirable that information transferred in the network cannot be traced back to the origin of the information or it
is impossible to find the receivers of the information. Another question is, if it is possible to censor the information communicated or 
turn off the system from a single point in the system. These features are in many ways in contradiction with monitoring and logging
features presented in Section \ref{logging}. The most well-known system built to solve these problems is Freenet \cite{FREENET}. 
Freenet is not discussed here in more detail because it isn't scalable and can lose unpopular data. 
Moreover, anonymity has not been a goal in designing HCA.

\subsubsection{Access Control}

HCA provides storage for shared data and a service for locating of services in a location independent way. These functions require persistent
data to be managed by HCA, which requires resources. Shared data and services also need to be protected from identity theft type attacks
and basic access control is needed for both reading and writing of given information. Sometimes, publishers
want to restrict the set of clients able to read the shared information. It should also be possible to efficiently change and manage this subset of 
clients.

Secure implementation of access rights and scalability are difficult to combine in an unreliable network, because scalability is achieved
by caching and redirecting information in the communication nodes in between sender and receiver. We need to trust all these
intermediate components or to provide end-to-end security by some other means. If some client does not trust the HCA internal security
mechanisms, it is possible to use cryptographic methods to autheticate and encrypt the communicated data at the end points of the
communication.

For maximum security, HCA supports also local services that are not visible and accessible outside the given domain. It is possible, for example,
to restrict some data or service only to company intranet and protect the intranet with firewalls. These services are used identically to
global services by clients inside the domain.

\subsubsection{Monitoring and Logging}
\label{logging}

Monitoring network activity can be useful for testing and, for example, tuning configuration parameters and spotting problems.
Logging information about network usage can be used to provide additional security, accountability, testing, billing of users, and
accumulating statistics. To a certain extent these features can be implementation details, but sometimes it would be beneficial if 
the interface of the system supported, for example, service providers in gathering usage statistics of information provided by them. 
These features are, however, not implemented in the first version of the HCA.

\subsubsection{Naming, Description, and Location of Data}

The information shared in the system and addresses of services must somehow be described, identified, and found in the system. Because
location independent access of data and services has been a design goal of the HCA, we need at least some sort of global identifiers that
unambiguously denote a certain logical data or service. The implementation can then transparently map these identifiers to the current
location of the data or service in the network. 

We decided not to support meta-data descriptions and sophisticated trading of services by their
properties or even human-readable naming of services in the HCA, because these can be implemented on top of HCA. The support for 
global human-readable names would pose difficult problems, because names can refer to external world and, therefore, would require a central
authority to control the namespace. Also, the requirements for naming service and location service can be different: For example,
it would be reasonably safe to assume that name to logical identifier mapping does not require real-time updates but logical identifier to
location mapping does, for example, in the case of migrating objects.

\section{Other Approaches}

Because of the general nature of the topic and because distributed hash tables and scalable file sharing are currently a hot topic, 
there are numerous other systems
with similar characteristics, including streaming networks, content delivery networks, peer-to-peer architectures, distributed file systems,
co-operative caches, file sharing networks, distribution research platforms, parallel computing platforms, middlewares,
event delivery systems, and programming language distribution primitives. 
It would be a gargantuan undertaking to discuss even a significant fraction of these. Therefore, only
a representative selection of technologies have been chosen here to provide a overall view of the field and
give different reference points to compare HCA to. These technologies are briefly introduced here and compared to HCA in
Chapter \ref{ch:discussion}.

\subsection{BitTorrent}

BitTorrent \cite{BITTORRENT} is an already popular system for fast file sharing. Its
main feature is that it integrates well with WWW and is easy to use. BitTorrent does not use any meta-data
information for files and does not provide clients with a global search of files as many other more centrally managed file
sharing services do. Users must find files they want to download by some external means and there already exists services for finding
BitTorrent files.

To users BitTorrent is straightforward: After installing their downloader software, BitTorrent can be started by clicking on a hyperlink
to the file user wishes to download. After this, the file is being downloaded from other peer clients downloading the same file. At the same time
the downloader also uploads the file to other peers. This way the upload capacity of the original provider of the file is spared because
uploading is distributed to all downloaders. To publish a file for download through BitTorrent, a .torrent file describing file information
must be created and put on some web server. This file contains file name and length, hashing information of the file, and the URL of the associated
\emph{tracker}. Tracker is a component that communicates information about currently downloading peers to other peers downloading the same file 
so that they can
find each other. Tracker also collects some basic statistics of file usage, but otherwise it doesn't take part in distribution of the file.
Tracker uses a light-weight protocol over HTTP. Also a "seed" downloader or some of these that have the entire 
file must be provided and connected to the tracker of the file.

With BitTorrent, the publisher makes the decision to use BitTorrent for distribution of the file to save his own upload bandwidth. 
Downloaders must then use BitTorrent clients to get the file. Downloading peers try to reach fairness in usage of network resources
with a simple "tit-for-tat" algorithm and try to balance the download and upload bandwidth for the peer by providing as much upload bandwidth for
each peer that it is receiving download bandwidth. The system has worked well in practice even when there are some known techniques
to cheat in the BitTorrent protocol \cite{BITTORRENTCHEAT}. 

A tracker gives each peer a random list of other peers and a random graph of peers is formed. This kind of graphs are robust and stay
well connected in case of network failures. However, this also means that the locality of peers is not exploited. Because of this the system has
a long latency but because the system is intended mainly for distribution of large files and the bandwidth is the main
concern, this is not a serious problem. 
Files are also cut into fixed size pieces, typically a quarter of megabyte in length and hashes of these blocks are used to check
data integrity in peers. Peers can then freely try to download these blocks from all other peers they are connected to. The algorithm for
choosing the order of download for blocks affects the performance of the system and BitTorrent downloaders use random block for the first
download and then try to download "the rarest first". When almost all blocks have been loaded, a different "endgame model"
is used. Peers are allowed to use choking of upload to certain other peers if proportional bandwidth of download is not received.

Because for each file there must be an associated tracker that is contacted by every peer that wants to download the file, the tracker forms
a bottleneck for scalability and failures but this is not a very serious drawback in the normal use case of hundreds of simultaneous downloaders
because the overhead bandwidth used in communication with tracker is only about a thousandth part of the file transfer bandwidth and can
be optimized further.
Cohen \cite{BITTORRENT} reports that the largest known deployments have had over a thousand simultaneous downloaders.

BitTorrent is a good example of a system with a feasible business model: It integrates with a web server and 
needs very little effort from
the file publisher to share the file. This way the publisher can reach a much larger number of downloaders because the
total downloading bandwidth of the clients added together doesn't need to correspond with the upload bandwidth of the publisher as it does with
the HTTP protocol and a typical web server. After the data can be downloaded only using BitTorrent, clients are forced to use BitTorrent, too.
This, a certain amount of fairness provided by the BitTorrent algorithm, and its easy user interface, 
has made BitTorrent probably the most popular file sharing technology used in the Internet today \cite{POPULARITY}.

\subsection{Chord}

A peer-to-peer network is a system of peer nodes that each can function as a server or a client in communication. In a pure peer-to-peer
system there is no central server, but all peers have, more or less, equal responsibilities.

Chord \cite{CHORD} is a distributed peer-to-peer lookup protocol for building scalable and robust distributed software. 
It is developed in the Chord project in MIT. Chord provides only one operation:
for a given key, it efficiently maps the key onto a node. It is easy to implement a basic distributed hash table (DHT) which maps
keys to values with Chord: the value
associated with a key is just stored in the node which Chord returns for the key. Therefore, Chord can be called a DHT implementation.
This single operation implemented by Chord is a fundamental building block in peer-to-peer applications. Several distributed systems have been
built on top of Chord: for example, it has been used as a basis for the Herodotus web archival system by Timo Burkard \cite{HERODOTUS} and 
the Cooperative File System (CFS) developed by Frank Dabek \cite{CFS}. 

Chord nodes form a totally decentralized architecture and each node functions symmetrically. Nodes can also join and leave network efficiently
and the system can continue answering queries simultaneously. The Chord protocol is also scalable: as the number of nodes increases, communication cost,
latency, and per node state each increase logarithmically. Its lookup mechanism is provably robust even when facing frequent node failures and
re-joins. For each query, $O(log(n))$ nodes are visited as the key is routed toward the destination in the network of $n$ nodes. 
Each node must contain information on about $O(log(n))$ other nodes in order to route queries efficiently.

The original Chord system as presented in \cite{CHORD} did not exploit locality and the search path through nodes could 
move backward and forward geographically. In later versions this has been enhanced along the development of CFS. 
The basic Chord system does not incorporate access rights and
security mechanisms, which practically means that Chord is used as a building block for an enclosing peer-to-peer system and not as
a stand-alone lookup system. Among DHT implementations Chord has quite a simple architecture with
essential features of scalability and fault-tolerance.

\subsection{IP Multicast}

The \emph{Internet protocol} (IP) \cite{IP} resides in the network layer of the ISO OSI model and is a lower level techonology than others presented here. 
It is relevant because IP supports \emph{multicast messages} that can be used as an implementation technology for efficient sharing of data.

\newnot{TTL}
Both IPv4, which is the original protocol presented in \cite{IP} and its successor IPv6 \cite{IPv6}
multicasts can be used for sending messages to multiple receivers in an IP network. 
Each multicast message is sent to a multicast group address
which doesn't denote any physical address in the network. Everyone can send messages to the group addresses and it is not necessary
to be a member of that group. The IP multicast routing protocol uses the Time To Live (TTL) field of IP datagrams to limit
the distance from a sending host a given packet should be forwarded. To receive multicast packets, an application must join to a
multicast group it wishes to listen.

Multicast can be used to save network resources, when many destinations are needed. If the same information would be transferred by sending
replicated unicast messages separately to each destination, network bandwidth would be wasted in sending multiple copies of the same data.
For example, Ethernet provides support for multicast and broadcast at the hardware level and IP multicast protocol
can take advantage of this in local sub-networks for a significant improvement in performance. 
Technologies at higher layers cannot exploit this optimization directly, but they can use IP multicast for their implementation. IP routers
use different algorithms for internetwork routing of multicast messages: Routers can, for example, form spanning trees between subnetworks
so that loops in the network are avoided. Also, subnetworks without multicast group members can be ignored because they signal routers of
having no group members. It is possible to design these routing algorithms and network topology scalable, for example, by using the hierarchical 
structure of regions containing groups of subnetworks.

\newnot{MTP}
IP multicast is unreliable which means that there is no guarantee that all messages reach all recipients. For some applications this is
acceptable, for example, real-time streaming of video can lose parts of the stream if better performance and small latency is attained this way.
Some reliable multicast protocols have been developed on top of IP. For example Multicast Transport Protocol (MTP) \cite{MTP} provides atomicity and 
reliability for applications.

In practise, many Internet routers connecting sub-networks do not support multicast messages. This problem has been circumvented by MBONE
virtual network \cite{MBONE} that provides an Internet multicast backbone tunneled through IP unicast packets and connecting the islands of subnetworks
that support multicast.
 
\subsection{Globe Location Service}

\newnot{GLOBE}
\newnot{DSO}
The GLOBE project \cite{Globe} researches construction of large-scale wide area
distributed systems using a unifying paradigm called \emph{distributed shared objects} (DSO). They have produced an architecture with the
same name Globe, a Global Object Based Environment. The Globe architecture
shares many of the overall goals with our project, which makes it an interesting system to compare with.

The basic idea is that Globe distributed objects are completely autonomous with respect to properties like distribution, consistency, and 
replication of state, which provides object-specific solutions to each of these aspects. Distributed shared objects are logical entities 
that can span many address spaces in multiple computers, unlike in typical object-based middlewares which encapsulate objects inside 
servers. In each address space the DSO is represented by a local representative which is composed of several subobjects. The exact
composition of representatives may vary but usually semantic, communication, replication, and control subobjects can be found. The
subobjects handle the respective aspects of the distribution of the DSO.

\newnot{GLS}
\newnot{OID}
Globe uses Globe location service (GLS) described in the PhD thesis of Gerco Ballintijn \cite{GLOBELOC} and in \cite{steen01achieving} to
locate objects in the network. Each DSO is identified by a worldwide unique object identifier (OID), which is location independent and 
points to the same logical object the whole lifetime of the object. GLS can be used to map OID to an actual location of the object, which
can be, for example, a network address, a port number, and, possibly, the replication and communication protocol that should be used to contact the
object. Typically, only local representatives of the object are registered in the GLS. Globe has also a separate name service for mapping
human-readable names to OIDs. When a client needs to use a DSO, it passes the OID of the DSO to the run-time system which queries the GLS with it. 
Usually, GLS returns a contact address for the nearest replica (or replicas) of the DSO, which is then used for loading and creating the local
representative of the DSO to the local address space.

Globe location service works on a worldwide scale and has a scalable design. GLS divides the Internet into a hierarchy of domains
to exploit locality. Domains at the bottom of the hierarchy represent local networks 
like an office network or a university campus network. These
leaf nodes are combined to form the next layer of domains and this process is applied recursively until, 
at the top, a root domain encompasses the whole Internet. Each domain is associated with a directory node. Each directory node keeps track of
the locations of DSOs in its domain. For each DSO, either the actual contact address is stored or a set of forwarding pointers, which point to
some child directory nodes of the domain, indicating that the contact address of the object can be found in the subtree starting from those
child domains. Typically, the contact addresses are stored in leaf directory nodes. The location service is also designed to support mobile
objects that can change location.

The hierarchical structure does not pose a scalability
problem for the directory nodes residing in the root and other high-level nodes, because these nodes are partitioned into one or more directory
subnodes in much the same way as in HCA. The subnodes can reside on different computers and they handle only a specific part of the object identifier 
space which is divided by a special hashing technique. The algorithm and
protocol used are further explained in \cite{GLOBELOCSCALE} and \cite{GLOBELOC}.

   \chapter{HYBRID COMMUNICATION ARCHITECTURE}
\label{ch:architecture} 

Now that we have defined the problem of the thesis, it is time to present the solution.
In this chapter we will go through the general level architecture of HCA. The architecture is composed of interfaces and components
implementing the interfaces.

\begin{figure} [h]
\centering
\centerline{\epsfxsize 0.8\linewidth \epsfbox{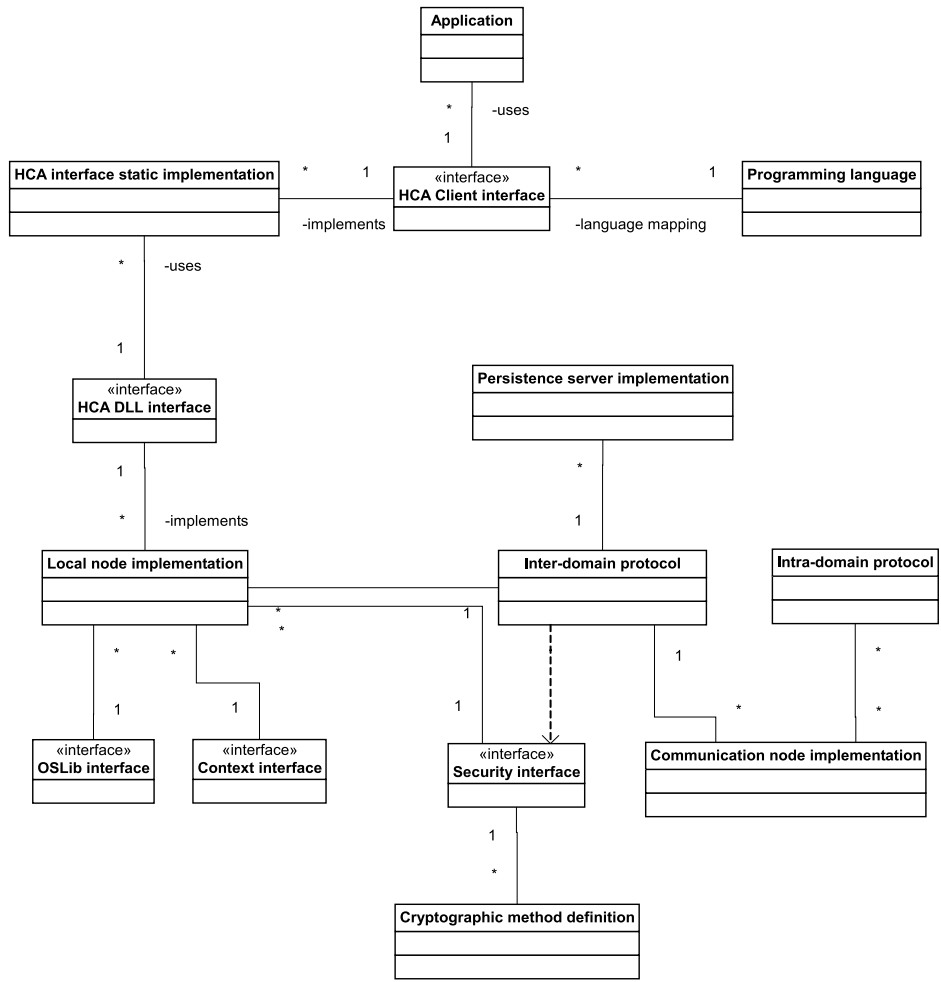}}
\caption{HCA static structure and dependencies}
\label{HCASS}
\end{figure}

Figure \ref{HCASS} depicts the static conceptual structure of HCA. The figure contains
the interfaces and protocols that form the architecture and components that will specialize the framework by implementing the
interfaces. These components together form an implementation of HCA. 

The following interfaces and protocols form the skeleton of the HCA framework:

\begin{itemize}
\item Client interface is used by applications on top of HCA,
\item DLL interface divides the client side implementation of HCA into dynamically and statically linked parts,
\item Context interface passes information from the execution environment to the application, 
\item Security interface provides an access to cryptographic functions,
\item OSLib interface provides an abstraction layer for different operating systems,
\item Inter-domain protocol is used by communication nodes of the HCA network,
\item Intra-domain protocol is used by communication nodes of the same domain as explained later.
\end{itemize}

The above components are described in more detail in the subsequent sections.

OSLib, Context interface and Security interface also have a language mapping for every supported
programming language. All the implementation components can have dependencies to these interfaces, but these dependencies are removed from
Figure \ref{HCASS} for clarity.

In a normal configuration, multiple applications with their HCA client interface static implementations and 
a single local node implementation (Figure \ref{HCASS}) are 
located on the same \emph{client machine} at the edge of the HCA overlay network. The client machine can be, for example, the user's
home computer or office workstation which is shut down every now and then. The local node communicates with the
nearest \emph{communication node} using the \emph{inter-domain protocol}. This node probably resides on a different machine typically administered by 
a service provider organization or it could be located in a company intranet server. 

The HCA overlay network is formed by many communication nodes 
structured in the form of hierarchical domain tree as will be explained in Chapter \ref{ch:domaintree}. The nodes are connected by the inter-domain
protocol and typically reside in servers that are continuously online.
Clients and \emph{persistence servers} can be thought of residing at the edge of the overlay network. 

Client interface, OSLib, \emph{Context interface} and \emph{Security interface} form together an operating-system- and hardware-independent
platform for components on top of HCA.

\section{Client Interface}

The HCA Client interface is the only part of the communication architecture visible to applications and higher-level layers of the middleware. 
Because HCA is intended to be programming language neutral, there are language mappings
of the client interface to each programming language supported.
The C++ client interface is presented in detail in Chapter \ref{ch:interfaces} and other language mappings can be
derived on the basis of it. In a nutshell, Client interface provides the interface for data sharing and message passing primitives and
access control.

The implementation of the client interface is normally divided to statically and dynamically linked parts so that it is possible
to change the implementation without recompiling applications. The statically compiled part is usually just a light-weight
portable wrapper for the DLL interface.

\section{DLL Interface and Local Node Implementation}

\newnot{DLL}
\newnot{API}
HCA DLL interface is an operating-system dependent application programming interface (API) for wrapping 
the local communication node to
a separate process running on the client machine. The DLL implementation shares the local communication node
between multiple client processes and implements the communication between client processes and the local communication node process using
fast inter-process communication, shared memory or other facilities provided by the operating system. 
The DLL interface is also independent of the programming languages
used for the programming of the clients. Because we use a dynamically linked API, it is possible to change the local implementation without
recompiling applications as long as API does not change.

Since a communication node is running on the client
machine, we will gain the benefit that no network traffic is needed for local communication through HCA between different client processes 
residing in the same machine. Also,
the nearest cache is implemented by the local node and it can store the cached information to a local hard disk. Therefore, often used
data can be accessed immediately without downloading it from network. The local communication node forms its own local domain in the
HCA overlay network. The significance of this will be explained in Chapters \ref{ch:domaintree} and \ref{ch:protocols}, 
but the most important consequence is
that some of the shared data and services contacted through HCA can be completely local so that other machines in the network cannot
access them. The local machine may also contain a number of persistence server processes attached to the local node. These two features
make it possible to use HCA as a general file system where local data and distributed data is accessed uniformly through the client interface.

The possible DLL interface specifications are outside the scope of this thesis, but generally the interface is
somewhere between a simplified procedural version of the client interface and a procedural version of the inter-domain protocol. 
The division of work between client processes and
local communication node process should be designed to maximize the performance of the system. If the statically linked part of
the implementation has lots of functionality, the DLL interface can be wrapped inside another operating system independent wrapper
for portability.

\section{OSLib}

\newnot{HAL}
OSLib programming interface provides a system-wide \emph{hardware abstraction layer} (HAL) for preventing
the components of HCA of having unnecessary dependencies to particular operating system or hardware.
With this modularization in implementations we gain the easy portability of HCA implementations over different
platforms. 

OSLib provides basic portable data types and interfaces for process, thread, concurrency control, interprocess communications,
date and time, timer, and local persistent storage.
In particular, it contains the abstraction of underlying network technologies. This is important because
the network is considered a more fundamental resource than other peripherals in distributed systems.
The basic idea is that other physical resources can be accessed through the network and this way we will
have transparent access to resources independent of their location. OSLib supports multiple network technologies
at the same time if the machine has access to several different networks. The basic primitives offered by OSLib for
communication are reliable network connection abstraction and unreliable connectionless datagram based primitive used for uni- and multicasting. 
They are intended to be straightforward match to TCP \cite{TCP}, UDP \cite{UDP} and IP multicast \cite{IP} protocols. 
The connection primitive is used for persistent communication between two nodes and datagrams are used for more 
light-weight temporary communication. The connection primitive can do some optimizations based on the
assumption on the typically long lifetime of connections.

HCA uses the connection primitive for implementing Inter-domain protocol and datagrams for 
so-called shortcut connections, which are explained in more detail in Chapter \ref{ch:protocols}. Inter-domain protocol can sometimes
also use multicast datagrams for optimization.

The full specification of OSLib is outside the scope this thesis.

\section{Context Interface}

Context interface is intended to pass information from the execution platform to the program. For example,
communication node implementations can use it to read the configuration information of the node. Client programs
can receive initial references to distributed resources from some external mechanism not visible to the client
implementation. One could say that context information is the bridge between program and its local environment.

The implementation mechanisms for passing context information can vary. It could be implemented as a process specific 
configuration file or as operating system environment variables.

The context interface is basically an immutable map from names represented as unicode strings to
values. The names can be arbitrary long so that it is possible to use global identifiers to name certain type of
information. The values of these context variables are untyped binary data that can be used for example to
deserialize objects. By deserialization we mean the process of translating binary data representing an object state back into an object.
The type checking of the data must be implemented by the application.

The full specification of the context interface is outside the scope this thesis.

\section{Security Interface}

Security interface contains a uniform interface for typical cryptographic algorithms. An abstract interface
for both public key and secret key methods are specified. Methods can be configured through this interface and their
basic properties can be examined. Security interface also specifies a SecurityManager class that provides an access
to all locally available methods. Also, an interface for a cryptographically strong random generator is specified. 
Public key methods should support generation of new public/secret key pairs, authentication of data by signing and de/encryption
of data. Secret key methods should support generation of new secret keys and de/encryption of data.

Both public and secret key methods are needed for securing HCA communication effectively. Public key methods
are used for authentication of nodes and en/decrypting of the initial handshake protocol where a secret key is exchanged 
between communicating nodes for en/decryption of further communication. Secret key methods are better suited for
actual encryption of communications because they are usually significantly faster than public key methods.

Because new cryptographic algorithms are constantly developed, old ones are compromized, and increasing available
processing power changes the underlying assumptions about needed key lenghts, it is necessary to have
an API that supports dynamic installation of new methods. This can be implemented by specifying a DLL interface
for cryptographic methods and implementing a local repository of available methods. This way new methods can
be added to the system without recompilation while the programs are running. The cryptographic methods have standard names 
under the security interface
and all implementations under a given name must work together so that encrypted network packets can
be understood at the both ends of a connection.

The full specification of security interface is outside the scope this thesis.

\section{Inter-domain Protocol}

Inter-domain protocol is used between all communication nodes to form the overlay network of HCA and it is explained in detail 
in Chapter \ref{ch:protocols}.
Cryptographic methods provided by the security interface are used to secure the protocol over an untrusted network and, therefore,
we have a dependency between Inter-domain protocol and Security interface. Nodes are locally configured with public key identities
of some other nodes they initially connect to join the network.

\section{Intra-domain Protocol}

The overlay network is partitioned into domains as explained in more detail in Chapter \ref{ch:domaintree}. Each domain
can use their own intra-domain protocol for coordination of the nodes inside the domain. Nodes must implement the intra-domain
protocol used by their enclosing domain. Small domains do not necessarily need an intra-domain protocol for managing nodes because it can be
achieved by configuring each node manually. 

Some implementation possibilities for intra-domain architectures and protocols are discussed in Chapter \ref{ch:protocols} but
a complete specification of such system is outside the scope of this thesis.

\section{Implementation}

A complete implementation of HCA must have at least the following components:

\begin{itemize}
\item Client interface static implementation
\item Local node implementation
\item OSLib implementation
\item Context interface implementation
\item Security interface implementation
\item Communication node implementation
\item Persistence server implementation
\end{itemize}

The components can be developed separately because they are separated by well defined interfaces. For example, different 
implementations of communication node can be mixed together. This can be useful because nodes residing in different parts of the
HCA overlay network topology can benefit from a different set of optimizations. 

\subsection{Communication Node Implementation}

HCA communication nodes are identical building blocks which are used to build the HCA overlay network.
They need constant resources in a well-designed domain hierarchy and can, therefore, be
built from standard, cheap, bulk components. Typically, a node is implemented as a process with some dedicated disk space
for caching and it is run on a network server. 
The inter-domain protocol is optimized to work efficiently with nodes that have relatively high availability and are always turned on in
absence of failures.
Typically, only local communication nodes at the edge of the overlay network are run on volatile computers like home computers. 

Multiple node processes belonging to different domains can be hosted by a single network server and there can even be implementations
that can run multiple logical nodes inside a single process. In this sense, communication nodes can be called virtual or logical.

Individual nodes are not required to have access to replicated or fault-tolerant persistent storage so that they
are easy to implement and can be relatively easily added or removed from the HCA network. 
The basic function of a communication node is to route and cache information produced by clients but they 
also participate in the implementation of mapping location independent identifiers denoting logical data or service 
to their current locations.

The domain of a communication node determines the geographical or network topological scope of the node, i.e., what 
is the maximum distance of connections
it needs, what kind of underlying network is used for those connections, how many nodes are in the same domain, what intra-domain
protocol they use if any, and if multiple administrative organizations are involved in managing the domain. Different
nodes are run in different environments and, therefore, they can benefit from different optimizations. Thus, it is useful that multiple
implementations can be used to build a single HCA overlay network.

Communication nodes can also have implementation-specific logging, configuration, and monitoring features.

\subsection{Persistence Server Implementation}

The communication node semantics do not incorporate any critical persistent state and they can relatively easily be 
replaced. All persistent state in HCA is modularized to be a responsibility of persistence servers residing at the edge
of the HCA network. They are used for storing:

\begin{itemize}
\item 
replicas of shared data published by clients,
\item
data structures used for managing the location independent identities used for communication, and
\item
access right management information.
\end{itemize}

Persistence servers in the HCA architecture are numerous, possibly dummy storage bricks
that can be implemented, for example, as a plug-in to a local communication node implementation. 
Persistence servers can be unreliable and they do not have to provide protection from accidental loss of data themselves because multiple
persistence servers can be used to replicate the same logical data and inter-domain protocol is used to coordinate the replication.
For example, there could be a persistence server on user's computer attached to the local node using a local hard disk for storing the
information. The protocol used by persistence servers is explained in detail in Chapter \ref{ch:protocols}.

Persistence servers have simple access right and quota mechanisms for preventing unauthorized use of resources. 
   \chapter{CLIENT PROGRAMMING INTERFACE}
\label{ch:interfaces}

\sloppy
\newnot{IDL}
We will next present the C++ client programming interface, which is used by applications and higher-level layers
to access the HCA network.
HCA is designed to be programming language neutral but we have selected C++ to formalize the concepts of the interface.
It should be fairly straightforward to map this interface to other programming languages. We could have used some
language neutral \emph{interface definition language} (\emph{IDL}) that have standard language mappings, but 
the automated language mappings can produce clumsy interfaces that do not use the idioms and concepts of the target
language efficiently.

\begin{figure} [ht]
\centering
\centerline{\epsfxsize 0.85\linewidth \epsfbox{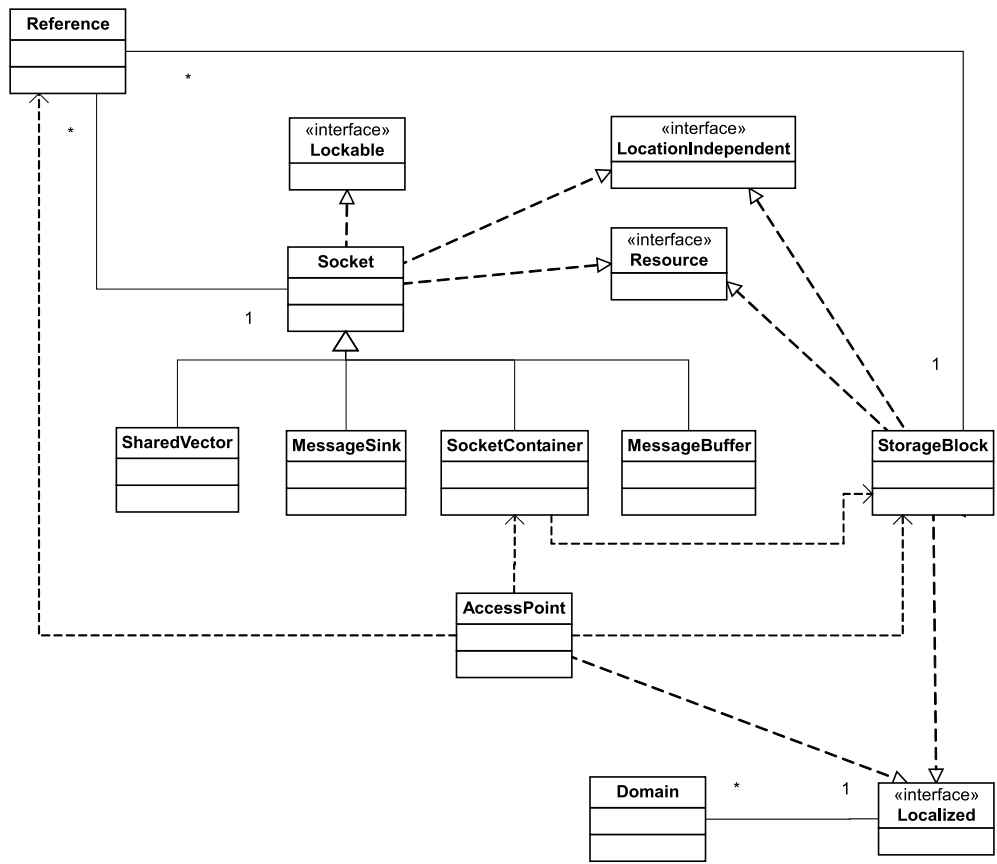}}
\caption{Static structure of the C++ HCA client interface}
\label{classdiag}
\end{figure}

The class diagram of the C++ client interface is depicted in Figure \ref{classdiag}. Exceptions, event
listener interfaces, and access right-related interfaces have been left out of the diagram for the clarity of the presentation.

The client interface introduces two basic communication concepts used by HCA: 

\begin{itemize}
\item
a class \verb1SharedVector1 for publishing and streaming data, and
\item 
a class \verb1MessageSink1 for creating a HCA socket that can be used for receiving messages.
\end{itemize}

Both primitives have
a common superclass \verb1Socket1. Clients are completely uncoupled from each other and communicate indirectly by using
HCA sockets. Each socket has a global, unique, location-independent identity which can be used to unambiguosly 
point to it from any access point
of the HCA network. Clients can possess socket references, each of which are used to refer to a certain socket. When a client creates
a new socket, it also attains a socket reference pointing to the created socket. It can then send the reference to other clients by
any channel and they can use the reference to communicate with the socket. Socket references point their whole life-time to the same
logical socket and if the socket is deleted, all references to it cease to function and never point to any socket. 

Each client uses each socket in a role of either a writer or a reader. For \verb1SharedVector1 the writer is a special client that 
can update the state of the shared data while readers can only observe the state of the vector. 
For \verb1MessageSink1 it is the opposite: readers
are special clients that can receive the messages sent by writers to the \verb1MessageSink1. 
There can be only one writer for \verb1SharedVector1 and one reader for \verb1MessageSink1 at any time.

\emph{Anycast} semantics of message passing means that a message is sent to only one of many possible receivers.
It is particularly useful for implementing a popular service under a single interface when clients of the service are 
independent of each other.
HCA does not support anycast semantics directly, because it can be implemented on top of HCA efficiently. For example, a service could use
a \verb1SharedVector1 for publishing references to all available \verb1MessageSink1 sockets for the service, 
their location in the overlay network and the load of
each replica. Clients can then choose an appropriate target based on a utility function of the distance to and load of the target. Because target
is chosen deterministically at client-side, it is possible to use a stateful protocol when communicating with the service because 
the same target of the anycast messages can be guaranteed. All this logic can be hidden behind a \emph{facade} \cite{PATTERNS} 
interface from the actual client code.

Sockets can be either persistent or their life-time
can be bound to the life-time of the client process that created them. Persistent sockets require storage resources which must be
granted to the socket when it is created. HCA provides also a simple access rights mechanism which can be used to restrict operations
performed by different clients. To implement access rights, clients are also identified in the system but
these client identities are never used for direct communication. 

\section{A General Description of Client Interface}

The whole client interface is \emph{thread-safe}, which means that the implementations of the classes handle locking of resources internally
and can be used safely with multiple threads.
The client interface uses asynchronous methods almost exclusively. These return their results in a callback to an object implementing
the corresponding listener interface. Asynchronous methods were used because network communication is required to perform the operation
and a blocking implementation would have wasted either threads or processor cycles. The listener interfaces for these asynchronous
interfaces are not presented in this thesis, but mostly they only have trivial callback methods that take the arguments described in the text.
The listener interfaces all have names that end with "Listener".

All C++ classes and interfaces presented in subsequent sections are located in \verb1hca1 namespace. Client interface has many
interfaces, which are implemented as C++ abstract classes.

\section{AccessPoint}

\verb1AccessPoint1 is a \emph{singleton} \cite{PATTERNS} object which functions as an entry point for Client interface. 
The singleton object can be accessed with the static member function \verb1instance()1.
\verb1AccessPoint1 provides an operation for creating new socket containers, 
which are used to manage persistent resources, socket policies,
and group sockets into hierarchical structures as will be explained in section \ref{containerSec}.

\verb1AccessPoint1 implements interface \verb1Localized1 and is associated with the current location of the client in the HCA domain tree. 
The location of the client
can change if the client is migrating. The interface \verb1Localized1 accepts \verb1DomainListener1 objects, 
which are notified of change in the enclosing
domain structure, as explained later in Chapter \ref{ch:domaintree}. Each domain has a set of \emph{boundaries}, which are represented as strings. 
Boundaries can be used to limit the visibility of sockets to a some enclosing domain in the overlay network. Each domain object also
has a method for requesting \verb1Identity1 of the domain, which can be used to authenticate the domain globally. 
Objects of the class \verb1Domain1
are immutable and the method \verb1getDomainHierarchy()1 of the interface \verb1Localized1 must be called again 
if the domain structure has been updated.
Root domain can be found always at the index 0 of the vector returned by \verb1getDomainHierarchy()1 
and the most local domain is at the last index of the vector.

The interface for the class \verb1AccessPoint1 is given below:

\begin{mylisting}
\begin{verbatim}
template <class T>
   class EventSource {
      virtual void addListener(T*) = 0;
      virtual void removeListener(T*);
      virtual void removeAllListeners();
  };


class Domain {
public:
   string getName();
   Identity getIdentity();
   vector<string> getBoundaries();
};


class Localized : public EventSource<DomainListener> {
   virtual vector<Domain> getDomainHierarchy() = 0;
};


class AccessPoint : public Localized {
public:
   void createSocketContainer(vector<Principal> &useRights, 
                              string &name, Identity &initialOwner, vector<StorageBlock*>&, 
                              uint32 minReplicas, uint32 maxReplicas, vector<string> &boundaries, 
                              ContainerFactoryListener *);
   void commitSocketContainer(Reference&, ContainerFactoryListener *);

   vector<Identity> &getCertificateAuthorities();
   void setCertificateAuthorities(vector<Identity>&);

   static AccessPoint *instance();
protected:
   AccessPoint();
   AccessPoint(const AccessPoint&);
   AccessPoint& operator=(const AccessPoint&);
   ~AccessPoint();
};
\end{verbatim}
\end{mylisting}

The asynchronous method \verb1createSocketContainer()1 is used for creating new socket containers. 
It is given a vector of pointers to \verb1StorageBlock1 objects,
which represent persistence servers and are used for replicated storing of the created container. 
Each \verb1StorageBlock1 object implements a
interface \verb1Resource1, which can be used to select which user rights are used for accessing the \verb1StorageBlock1.
If the vector is empty, the created
container will be temporary. This means that it is stored locally, 
can have a maximum of one replica, and the lifetime of the container and all its contents are bound
to the lifetime of the client. \verb1initialOwner1 will initially be granted all access rights to the container.
It can be different than the \verb1Principal1 used for each \verb1StorageBlock1. 
The argument \verb1useRights1 is used to determine the set of principals
whose access rights are used in performing the operation. 
The argument \verb1minReplicas1 determines the minimum number of working replicas of the data required for normal operation
of the socket to continue. The argument \verb1maxReplicas1 is used to set the ideal number of working replicas HCA tries to keep up 
when enough resources are available. The process of creating a new socket container is two-phased: First, a callback to the
provided \verb1ContainerFactoryListener1 object returns a preliminary reference to the created container. The client can then store
the preliminary reference persistently, and after that, call asynchronous method  \verb1commitSocketContainer1. In the second callback
the client gets certification that the container is created persistently. Both of these phases can fail and client is notified of this
to the \verb1ContainerFactoryListener1 callback. The two-phase protocol for creation of the containers prevents a creation of ghost
containers that do not have any references pointing to them in the case of a client failure.

Methods \verb1getCertificateAuthorities()1 and \verb1setCertificateAuthorities()1 can be used to query and change, respectively, the set of
authorities trusted on general level (in contrast to per reference basis) by the access point for authenticating sockets and persistence servers.

\section{Location Independent References}

Clients can access a given socket by using a socket reference which they have acquired by any means. They may have
either stored the reference locally, they could receive it through some other HCA socket, or by some external means. 
Each reference points to
a single logical socket the whole lifetime of the reference. The socket can be removed, which means that some
references can point to unexisting sockets. This kind of socket references are called \emph{dangling}. 
This is not a problem since when client tries to access the socket using
the reference, an error message of non-existing socket is returned. After the socket has been removed, the reference
cannot ever point to any other socket. It is always possible for the client to distinguish a dangling pointer from a connection error.

The method \verb1open()1 is used for opening a proxy object, which is used to handle communications to the resource 
referenced by a \verb1Reference1 object.
If the referred entity does not exist, an error is returned. The method \verb1open()1 is asynchronous and returns to a call-back 
either a dereferenced location-independent entity 
or a failure to the provided factory object. References are basically used to refer to different socket types and 
\verb1StorageBlock1s, which represent persistence servers. The classes \verb1Socket1 and \verb1StorageBlock1 both implement 
the interface \verb1LocationIndependent1.
Clients can dynamically cast \verb1LocationIndependent1 objects to their actual types.

The method \verb1equals()1 returns true if and only if the argument points to the same logical entity as the \verb1Reference1.
The method \verb1getId()1 returns a globally unique integer that identifies the socket or persistence server pointed by the reference.
method \verb1getCertificateAuthorities()1 returns a vector of accepted certificate authority identities provided by the reference. If the
vector is empty, Client interface implementation may still trust some general authorities and the secure authentication of
the entity pointed by the reference can be done. The method \verb1setCertificateAuthorities()1 can be used to change the trusted authorities.

The interface for \verb1Reference1 is given below:

\begin{mylisting}
\begin{verbatim}
class Reference : public Serializable {
   virtual Integer &getId();
   virtual vector<Identity> &getCertificateAuthorities();
   virtual void setCertificateAuthorities(vector<Identity>&);

   virtual void isDangling(ReferenceListener *);
   virtual void open(vector<Principal> *useRights, LocationIndependentFactoryListener *);
   virtual bool equals(Reference&);
};
\end{verbatim}
\end{mylisting}

\section{Socket}

HCA \verb1Socket1 is the abstract base class for communication primitives offered by HCA. 
The classes \verb1SharedVector1, \verb1MessageSink1, \verb1SocketContainer1, \verb1MessageBuffer1,
\verb1Group1, \verb1AccessRight1, and \verb1Role1 are derived from \verb1Socket1. 
The class \verb1Socket1 implements the interfaces \verb1LocationIndependent1, \verb1Lockable1, and \verb1Resource1.
All sockets have a HCA wide unambiguous identity and can be referenced using location independent global references, 
which are implemented by the class \verb1Reference1. 

Sockets are always contained inside socket containers, which are represented by the class \verb1SocketContainer1.
Socket containers can be used to set common policies to all sockets they contain. All sockets require storage resources and
they consume resources provided by their container. If a container is associated with some persistence
servers, the lifetime of the sockets it contains are independent of the lifetime of the clients that created the sockets. Clients can migrate
in the HCA network and still keep using the same sockets.

\begin{mylisting}
\begin{verbatim}
class Lockable : public EventSource<LockListener> {
   virtual string *getClientId();
   virtual void forceAcquireLock(string &clientId, LockListener *) = 0;
   virtual void acquireIfNotLocked(string &clientId, LockListener *);
   virtual void waitForLock(string &clientId, uint32 timeInMs, LockListener *);
   virtual void releaseLock(LockListener *);
   virtual bool hasLock();
};
\end{verbatim}
\end{mylisting}

The interface \verb1Lockable1 can be used for locking and unlocking the socket. 
The method \verb1forceAcquireLock()1 always succeeds in locking the socket in the
absence of network errors. The method \verb1acquireIfNotLocked()1 locks the socket if there is no previous 
lock of the socket by some other client. The method \verb1waitForLock()1
can be used to set a maximum time the lock is tried to get before giving up. If the time parameter is 0, an arbitrary time can be waited for
acquiring the lock. The method \verb1hasLock()1 can be used to check if the proxy object has the lock and \verb1releaseLock()1 
can be used to release the lock.
If the proxy object is deleted or in the case of long network error, the lock can be released automatically. Most of the operations to
sockets can be done without first locking the socket, but some operations that change the state of the socket require the lock to the socket
to prevent multiple clients simultaneously changing the socket. All methods of interface \verb1Lockable1 except \verb1hasLock()1 
are asynchronous, which
means that the methods return immediately but signal all attached \verb1LockListener1 objects with a call-back of the result when it arrives. 
Attached listeners
are also signaled of change in the lock state of the socket. The argument \verb1LockListener1 of asynchronous methods can be NULL. 
In that case only listeners attached to interface \verb1Lockable1 are signaled.

\subsubsection{Socket Scope}

Clients of the HCA reside at every moment in some branch of the domain tree. If the domain hierarchy is well designed,
it will reflect the physical architecture of the underlying network, as well as administrative, social, and geographical
boundaries. The hierarchy can be taken advantage of clients by setting a scope for each socket created. 
This is useful for achieving the maximum security for sockets which are not used globally. In essence, the communication nodes
implement firewalls that prevent the use and limit the visibility of the sockets outside the scope of each socket.
Every socket can be associated with multiple boundaries using interface \verb1LocationIndependent1. 
Boundary values are represented as strings. 

Each domain can be configured
by the administrators of the domain to have a set of boundaries associated with the domain. HCA guarantees that
sockets associated with a given boundary denoted by a string are never visible to domains outside the smallest domain having the configured
boundary and which is enclosing the persistence servers used for storing the socket. 
If a scoped socket is persistent, all storage resources allocated for the socket must
reside under the smallest domain containing the boundary. If socket is not associated with any boundary, it is by default a global socket and
visible all the way to the root domain.
However, if the client wishes to limit the visibility of the socket to some
local domain, it is possible. This is useful if the client needs absolute certainty of the security of the socket and
no global access to the socket is needed. If a socket is associated with a boundary which is not found in any enclosing domain, the socket
is not visible at all.

A list of standard boundary strings, which should be used when applicable by the implementors of the overlay network 
is starting from local to global:

\begin{enumerate}
\item Local machine
\item Local network
\item Intranet
\item Service provider
\item Country
\item Continent
\item Root
\end{enumerate}

For example, the interprocess communication should be implemented with temporary socket inside the "Local machine"
boundary. Sensitive company data can be stored behind sockets in the domain "Intranet" that is protected by a firewall.

It should be noted that the above hierarchy is divided, on purpose, on boundaries usually separating networks of different
speed. Interprocess communication can be have couple order of magnitudes greater bandwidth and smaller latency
than a local area network, which, on the other hand, can be an order of magnitude faster than access to an Internet service provider. When
the HCA overlay network is designed this way and the data is fetched from the most local domain having the data cached,
the use of slower outside network can be completely avoided.

If boundaries associated with the socket are changed, all attached \verb1LocationIndependentListener1 objects are signaled.

The interface \verb1Socket1 is presented below:

\begin{mylisting}
\begin{verbatim}
class Socket : public LocationIndependent, public Lockable, public Resource, 
               public EventSource<SocketListener> {
public:
   virtual void openContainer(LocationIndependentFactoryListener *) = 0;
   virtual Reference *getContainerReference();
   virtual void destroySocket(SocketListener *);
protected:
   Socket();
   Socket(const Socket&);
   Socket& operator=(const Socket&);
   virtual ~Socket();
};
\end{verbatim}
\end{mylisting}

The method \verb1getReference()1 of the interface \verb1LocationIndependent1 returns a new socket reference, which refers to the socket.
The asynchronous method \verb1generateNewIdentity1 can be used to produce a new public key identity 
for the logical entity. This invalidates all existing references and certificates containing the old identity.  The process has two phases: 
First a callback to the provided \verb1LocationIndependentListener1 object is performed and a new reference to the entity returned.
The new reference is not yet valid but asynchronous \verb1commitNewIdentity()1 must be called and after the call returns in callback,
the identity is changed permanently. The client can store the new reference persistently before calling \verb1commitNewIdentity()1
and this way the possibility of ghost sockets is prevented if client crashes during the creation of the new identity.
A new working reference can be acquired with the method \verb1getReference()1. The method \verb1getCertificates()1 returns all certificates of
authenticity the \verb1LocationIndependent1 entity provides at the moment. Asynchronous method \verb1setCertificates()1 can be used to
change the set of certificates.

\begin{mylisting}
\begin{verbatim}
class Certificate : public Serializable {
   virtual Identity &getAuthority();
   virtual Integer &getId();
   virtual Identity &getIdentity();
   virtual Integer &getVersion();
   virtual Date *getStartDate();
   virtual Date *getEndDate();
   virtual string &getAuthorityName();
   virtual string &getDescription();

   Certificate(Principal &Authority, string &authorityName,
               string &description,
               Integer &id, Identity&, Date *start, Date *end);
   virtual ~Certificate();
};
\end{verbatim}
\end{mylisting}

\verb1Certificate1 objects encapsulate a signed certificate of authenticity of an entity in the HCA network. Each certificate is signed
by an authority that is identified by its public key. Each certificate contains the public key identity of the authenticated entity,
an integer id identifying the entity, textual name of the signing authority, version number of the certificate, 
textual description of the entity and possible start, and end date for the validity of the certificate.

\begin{mylisting}
\begin{verbatim}
class Resource : public EventSource<ResourceListener> {
   virtual void useRights(vector<Principal>*) = 0;
   virtual void addPrincipal(Principal&);
   virtual void removePrincipal(Principal&);
   virtual vector<Principal> listPrincipals();
   virtual vector<Role> getDefaultRoles();
   virtual vector<AccessRight> getAccessRights();
};
\end{verbatim}
\end{mylisting}

The interface \verb1Resource1 can be used to query access rights and default roles associated with the access control of the socket.
Each socket must have access rights named "lock", "force lock", "change boundaries", and "destroy socket", 
which give rights to lock the socket, force the lock of the socket, change boundaries
of the socket, and destroy the socket, respectively. Everybody is allowed to read the state of the socket. Each socket must has
default roles named "reader", "writer", and "owner". All principals granted the "owner" role should have all access rights to the socket.

If a client crashes when it has locked sockets, the sockets remain locked until some other client uses method \verb1forceAcquireLock()1 to 
attain the lock. The lock can be acquired also without forcing if it requested with the same client identification.

The methods \verb1useRights()1, \verb1addPrincipal()1, \verb1removePrincipal()1, and \verb1listPrincipals()1 are 
used for setting the principals whose rights are used 
to access the resource when performing operations provided by the object that implements the interface \verb1Resource1.

\begin{mylisting}
\begin{verbatim}
class LocationIndependent : public EventSource<LocationIndependentListener> {
   virtual Reference getReference() = 0;

   virtual Integer &getId();
   virtual string &getName(); 
   virtual Identity &getIdentity();
   virtual vector<Certificate> &getCertificates();
   virtual void setCertificates(vector<Certificate>&, 
                                LocationIndependentListener*);
   virtual void generateNewIdentity(LocationIndependentListener*);
   virtual void commitNewIdentity(LocationIndependentListener*);

   virtual vector<string> getBoundaries();
   virtual void setBoundary(string&, LocationIndependentListener*);
   virtual void removeBoundary(string&, LocationIndependentListener*);
   virtual void removeAllBoundaries(LocationIndependentListener*);
};
\end{verbatim}
\end{mylisting}

Each socket is given an immutable textual name or desription that can be queried with the method \verb1getName()1.
The method \verb1getIdentity()1 returns the public-key-based global identity of the socket,
\verb1getId()1 returns a globally unique integer that identifies the entity implementing interface \verb1LocationIndependent1. 
The method \verb1getContainerReference()1 returns a pointer to a reference to the container that contains the socket. 
If the socket is a root container, \verb1NULL1 is returned. If \verb1SocketContainer1 socket is moved, 
its parent container may change. This is 
notified to all attached \verb1SocketListener1 objects. The method \verb1openContainer()1 can be used to directly open the enclosing container by
using the same credentials as the socket is used with.
The method \verb1destroySocket()1 destroys
the socket permanently and releases all persistent storage allocated to it if it succeeds. The destructor only destroys the proxy
object, which is used for communication to the socket.

\section{SharedVector}

The class \verb1SharedVector1 is a primitive for sharing and streaming data.
The data is stored in a vector where each element of the vector contains variable length binary data. The elements are indexed 
from $0$ to $n-1$ where $n$ is the size of the vector. The initial size is $0$. These elements can be added or removed freely in the 
lifetime of the vector and the size of the vector changes accordingly. The addition of elements is not restricted to the end of 
the vector. If new elements are added in the middle of the vector, every index is increased after the added element. 

The class \verb1SharedVector1 is implemented as a sparse vector which means it can store efficiently vectors containing lots of empty elements. 
If a new element is added beyond the last index $k$ of the vector, it is assumed that all elements from $k+1$ to $n-1$ are empty elements, 
where $n$ is the index of the new element added. \verb1SharedVector1 can be used to store large amounts of data and it can be read in parts 
by loading only a small window of indices at a time.

There can be only one writer at a time for a \verb1SharedVector1. The \verb1SharedVector1 socket must be locked by the writer 
before its state can be changed.
HCA protocol only transfers changes to a state of a \verb1SharedVector1 and intermediate communication nodes 
can cache previous states of the
vector. Readers of a \verb1SharedVector1 can \emph{subscribe} to a range of indices of the \verb1SharedVector1 
and they will automatically receive
all updates to the selected indices. Readers are also notified of updates and it is guaranteed that as long as the connection stays on,
no intermediate states of the \verb1SharedVector1 are lost.
Therefore, \verb1SharedVector1 can be used for streaming data to multiple receivers. If the writer of a \verb1SharedVector1 is faster than
the readers, it is possible that some readers face a buffer overflow and must stop receiving the changes.

\subsubsection{Streaming states}

The lifetime of a \verb1SharedVector1 consists of numbered states that change over time. The writer of the \verb1SharedVector1 
can make arbitrary 
changes to the vector and when a commit is performed, a new public state for the \verb1SharedVector1 is generated. The readers of 
the vector can only see these committed states.

\begin{figure} [ht]
\centering
\centerline{\epsfxsize 0.6\linewidth \epsfbox{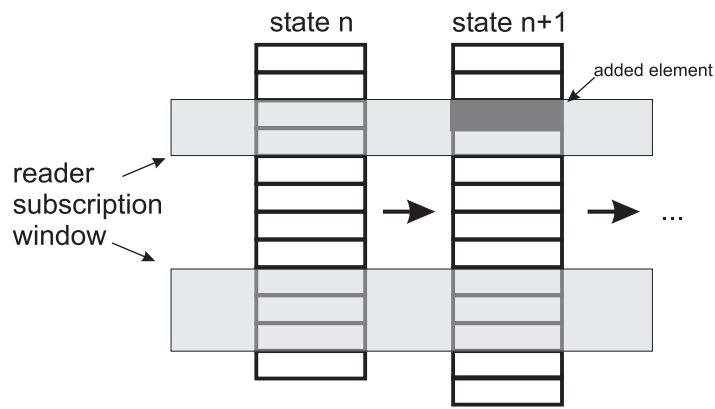}}
\caption{SharedVector update and reader subscription window}
\label{sharedvectorupdate}
\end{figure}

In Figure \ref{sharedvectorupdate}, a \verb1SharedVector1 reader subscription window and a stream of \verb1SharedVector1 states are depicted.
If a new state is generated by the writer of the \verb1SharedVector1, all readers are notified of the state change independent of
whether it affects entities inside their subscribed index range.

The interface \verb1SharedVector1 is given below:

\begin{mylisting}
\begin{verbatim}
class SharedVector : public Socket, public EventSource<UpdateListener>, 
                     public EventSource<ConnectionListener>,
                     public EventSource<SharedVectorListener> {
public:
   Integer size();
   Integer state(); 

   bool isAuthenticated();
   void setAuthenticated(bool);

   void add(Integer index, Integer size, uint8 *dataPtr) throw(NoLockException);
   void append(Integer size, uint8 *dataPtr) throw(NoLockException);
   void remove(Integer index) throw(NoLockException, OutOfBoundException);
   void removeAll() throw(NoLockException);
   void set(Integer index, Integer size, uint8 *dataPtr) throw(NoLockException);
   void setSize(Integer newSize) throw(NoLockException);
   void modify(Integer index, StateTransformer *tPtr) throw(NoLockException, OutOfBoundException);
	
   void commit(PersistenceListener*) throw(NoLockException);

   const uint8 *get(Integer index, Integer &size) throw(OutOfBoundException);

   void commitSubscriptionWindow();
	
   void subscribeAll();
   void unsubscribeAll();

   void subscribeRange(Integer sIndex, Integer eIndex);
   void unsubscribeRange(Integer sIndex, Integer eIndex);

   void snapshot();

   bool isConnected();
   void connect(ConnectionListener*); 
   void disconnect();

   Integer unconsumedStates();
   Integer unconsumedVolatileStates();

   void lastState(UpdateListener *) throw(NoLockException); 

   void nextState() throw(OutOfBoundException);
   void nextVolatileState() throw(OutOfBoundException);

   iterator getModifications(); 

   virtual ~SharedVector();

private:
   SharedVector();
   SharedVector(const SharedVector&);
   SharedVector& operator=(const SharedVector&);
};
\end{verbatim}
\end{mylisting}

The readers of the \verb1SharedVector1 can use it in two different modes: subscribing a window of indices of the vector and receiving 
state changes using push semantics or by requesting a snapshot of a window of indices of the \verb1SharedVector1. In Client interface
a reader sees a queue of locally loaded states and can consume them by calling \verb1nextState()1, \verb1nextVolatileState()1, 
or \verb1lastState()1.
If there are loaded states in the queue and \verb1nextState()1 is called, a new state is selected for active handling. The reader can then 
query elements of the active state of the SharedVector with method \verb1get()1. The elements are raw binary data.
\verb1nextVolatileState()1 also selects a new active state for
handling, but it does not necessarily wait for persistence servers to store the state permanently. This way the reader may gain
a smaller latency of state updates but in some rare occasions the state of the vector can rollback to a previous state if persistence
servers malfunction. If a reader uses the \verb1SharedVector1 in the push mode, all states are received in order as long as the client stays
connected to the vector. If a communication error occurs and the client cannot receive some state, the connection is broken and the client
is notified. The client can re-establish the communication by calling the method \verb1connect()1. If the vector is used in pull mode by calling
the method \verb1snapshot()1, each call loads a new state to the received queue if the state of the vector has changed. 
This way the client is not,
however, guaranteed to receive all the intermediate states of the vector.

The method \verb1lastState()1 is used by a writer of the \verb1SharedVector1 after it has acquired the lock of the \verb1SharedVector1 socket.
This is an asynchronous method that will load the latest state of the vector for active handling. After the last state has been
loaded, the writer can use method \verb1modify()1 to modify the current state of the vector to produce a new state. Writers can also add and
replace elements to the vector without first loading the previous state of the vector. Writer can publish a new state of 
the \verb1SharedVector1
by calling the method \verb1commit()1. Writer is notified when the new state has been stored by the persistent storages.

Both writers and readers can use methods \verb1subscribeRange()1, \verb1unsubscribeRange()1, \verb1subscribeAll()1, and \verb1unsubscribeAll()1 
to change the window
of indices they are interested in. By calling \verb1commitSubscriptionWindow()1 the new window is requested. 
If the window has grown, all loaded
states are invalidated and the client must wait for the new elements to be loaded. If there are many old states in the queue, this
can cause the connection to break.

The method \verb1size()1 returns the number of elements in the active state of the vector. The method \verb1state()1 returns an integer 
state number of the active state.
The methods \verb1add()1, \verb1append()1, \verb1remove()1, \verb1removeAll()1, \verb1set()1, \verb1setSize()1, and \verb1modify()1 
can be used by the writer to make modifications to the active state of the vector.

The methods \verb1unconsumedStates()1 and \verb1unconsumedVolatileStates()1 return the number of loaded states waiting in the received queue.

The method \verb1getModifications()1 returns an iterator, which will iterate through changed indices of the active state from the previous loaded state.

\section{MessageSink}

The socket \verb1MessageSink1 is a communication primitive that enables clients to send messages to the reader of the \verb1MessageSink1.
A reader can start receiving messages by first acquiring the socket lock and then calling the method \verb1startReceiving()1. 
The reader is notified of new messages with a callback to the attached \verb1MessageListener1 objects. 
The interface \verb1MessageSink1 is given below:

\begin{mylisting}
\begin{verbatim}
class MessageSink : public Socket, public EventSource<MessageListener>, 
                    public EventSource<ConnectionListener> {
public:
   void setMaximumMessageLength(Integer, MessageListener *);

   void send(Integer size, uint8 *dataPtr, MessageBuffer *, Reference *fallbackAddress, 
             Integer maxTimeInMs, MessageListener *);

   bool isReceiving();
   void startReceiving() throw(NoLockException);
   void stopReceiving();

   Integer messageCount(); 
   Integer receiveNext(uint8* &dataPtr) throw(OutOfBoundException);
   void consumeNextMessage(MessageListener*);

   virtual ~MessageSink();

private:
   MessageSink();
   MessageSink(const MessageSink&);
   MessageSink& operator=(const MessageSink&);
};
\end{verbatim}
\end{mylisting}

The reader can set a maximum length for accepted messages using the asynchronous method \verb1setMaximumMessageLength()1 and 
communication nodes will filter out all messages longer than that. If a negative value is used, there is no limit for the size of
the messages.
The reader can stop receiving messages by calling the asynchronous method \verb1stopReceiving()1.
All received messages are stored in reader's receiving buffer. The number of messages in the receiving buffer can be queried with
the method \verb1messageCount()1 and the first message in the buffer can be read with the method \verb1receiveNext()1. The method \verb1receiveNext()1 
returns the length of the message in bytes and a pointer to the raw data of the message. The reader must eventually call 
the asynchronous method \verb1consumeNextMessage()1 before it can receive other messages. This sends a signal to the sending
\verb1MessageBuffer1 that message has been received and can be removed. \verb1MessageBuffer1 acknowledges the deletion of the message by
calling back the reader of the \verb1MessageSink1 and the given \verb1MessageListener1 object is notified of the deletion. 
After the callback, the reader can be sure to not to receive the same message again.

Writers can use the asynchronous method \verb1send()1 for sending messages to the \verb1MessageSink1 socket. A message is a block of raw 
binary data.
A writer can provide a \verb1MessageBuffer1 object as an argument to the method \verb1send()1. 
Objects of type \verb1MessageBuffer1 are also sockets and they can be
used for storing the messages persistently until they have reached the reader of the \verb1MessageSink1. 
By using \verb1MessageBuffer1, it is possible
to shut down the client that sent the message: \verb1MessageBuffer1 guarantees that the message is eventually delivered to the reader of 
the \verb1MessageSink1 if one turns up. The originator of the message is notified when the \verb1MessageBuffer1 
has stored the message for 
dispatching. It is also possible to set a maximum time for a message to reach its destination. A reference to another \verb1MessageSink1 can
be provided as an argument to the method \verb1send()1 and it is used as a return address for the message if it could not be sent to 
its primary destination in its designated time.

The interface \verb1MessageBuffer1 is given below:

\begin{mylisting}
\begin{verbatim}
class MessageBuffer : public Socket, public EventSource<MessageBufferListener> {
   Integer messageCount();
   Integer resourcesUsed();
   void clearAllMessages(MessageBufferListener *) throw (NoLockException);
   virtual ~MessageBuffer();
private:
   MessageBuffer();
   MessageBuffer(const MessageBuffer&);
   MessageBuffer& operator=(const MessageBuffer&);
};
\end{verbatim}
\end{mylisting}

The class \verb1MessageBuffer1 provides the method \verb1messageCount()1 for determining how many messages are currently stored in the buffer.
A method \verb1resourcesUsed()1 returns an integer denoting how much resources the messages in the buffer are consuming. 
All attached \verb1MessageBufferListener1 objects are notified if there is a change in these numbers. 
The contents of the \verb1MessageBuffer1 can be forcefully removed by calling the asynchronous method \verb1clearAllMessages()1.

\section{SocketContainer}
\label{containerSec}

Socket containers are used for arranging sockets into groups and setting policies for groups of sockets. 
Each socket is always associated with one or zero enclosing containers. 
\verb1SocketContainer1 is inherited from socket, which means that containers can be recursively nested.
Containers provide HCA with some means of structuring sockets and handling groups of sockets efficiently. 
It would be possible to
provide the same communication possibilities without containers, but a single huge flat address space of sockets would
be cumbersome to maintain without external bookkeeping. The socket container hierarchy is not designed for representing logical
relationship "containing" but to group sockets with similar communication policies like replication, access rights, scope, and resources.
Containers can be used also to abstract the distribution and the replication of data to multiple persistence servers.

Socket scope is limited by the scopes of its enclosing containers.

\begin{figure} [ht]
\centering
\centerline{\epsfxsize 0.7\linewidth\epsfbox{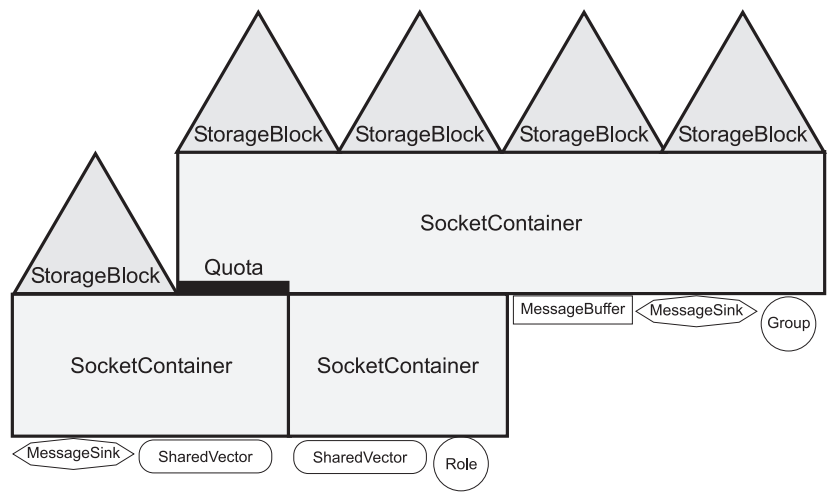}}
\caption{An example of socket hierarchy}
\label{container}
\end{figure}

Figure \ref{container} presents a possible tree of containers and sockets. The HCA address space consists of a forest of 
socket trees similar to the one shown in Figure \ref{container}. Each container has a set of storage blocks, which are used to provide
the persistent storage space for socket information and contents of the recursively contained sockets. 

The interface \verb1SocketContainer1 is presented below:

\begin{mylisting}
\begin{verbatim}
const int32 NOVALUE = -1;

class SocketContainer : public Socket, public EventSource<SocketContainerListener> {
public:
   int32 getMinReplicas();
   int32 getMaxReplicas();
   void setMinReplicas(int32, SocketContainerListener*); 
   void setMaxReplicas(int32, SocketContainerListener*);

   Integer resourcesAvailable();
   Integer resourcesUsed(); 

   void moveExistingContainerHere(SocketContainer&, SocketContainerListener*);

   vector<Reference> getStorageBlocks();
   void addStorageBlock(StorageBlock&, SocketContainerListener*);
   void removeStorageBlock(Reference&, SocketContainerListener*); 
   void changeStorageBlockUser(StorageBlock&);

   void setQuota(string, Integer, SocketContainerListener*); 

   void createSocket(string name, Integer quota, Identity *initialOwner,
                     SocketFactory&, ReferenceFactoryListener*);

   void openSocketByName(string name, LocationIndependentFactoryListener*);
   void getSocketReference(string name, SocketContainerListener*);
   void getAllSockets(SocketContainerListener*); 

   void removeSocket(string, SocketContainerListener*);
   void removeAllSockets(SocketContainerListener*);

   virtual ~SocketContainer();

private:
   SocketContainer();
   SocketContainer(const SocketContainer&);
   SocketContainer& operator=(const SocketContainer&);
};
\end{verbatim}
\end{mylisting}

If there are multiple storage blocks 
attached to a container, they can be used for both extending the available resources and providing replication of the persistent data.
Storage blocks attached to a parent container of a container are also used for storing the contents of the container. The usage of
resources of the parent containers by child sockets can be limited by setting a quota for the socket 
using the method \verb1setQuota()1 of the enclosing container. 
References to all attached persistence servers can be acquired with the method \verb1getStorageBlocks()1. 
The methods \verb1addStorageBlock()1 and \verb1removeStorageBlock()1 can be used to add and remove persistence servers
used for storing the container. The method \verb1changeStorageBlockUser()1 can be used to change the principals whose access rights are used 
to access a given persistence server.

Containers consist of multiple sockets that are identified by unique textual names inside the container. New sockets can be created
with the asynchronous method \verb1createSocket()1. The type of the created socket is determined by the provided \verb1SocketFactory1 object.
There must not already exist a socket with the same name in the container. The names of sockets can be maximum of 1000 characters long and
they may not contain character '/'.

Each container has two parameters to control the replication of their contents: \verb1minReplicas1 and \verb1maxReplicas1. 
\verb1maxReplicas1 is
the preferred number of replicas for the data: persistence servers should always try to maintain this number of replicas if there are
enough persistence servers and resources associated with the container. The parameter \verb1minReplicas1 sets the minimum number of 
working replicas
required for \verb1SharedVector1 sockets to signal their clients that the state is committed. These two parameters can be set by 
the asynchronous methods \verb1setMinReplicas()1
and \verb1setMaxReplicas()1 and their values queried by the methods \verb1getMinReplicas()1 and \verb1getMaxReplicas()1. 
Both parameters can be set
to have value \verb1NOVALUE1, which means that parameter values of the parent container are used. 
If root container uses \verb1NOVALUE1 setting,
the implementation of persistence servers may freely choose approriate values for the parameters.

The methods \verb1resourcesAvailable()1 and \verb1resourcesUsed()1 return integer values describing the amount of 
persistent storage space available for and used by the container.
The method \verb1moveExistingContainerHere()1 is an asynchronous operation for transfering 
a whole container and its contents under the current container. This can mean moving a large amount of data over the network.
This does not affect references pointing to the container and
the way clients access the container and its contents.

The asynchronous methods \verb1openSocketByName()1, \verb1getSocketReference()1, and \verb1getAllSockets()1 are for 
accessing sockets in the container.
When opening a socket by name using \verb1openSocketByName()1 or \verb1getSocketReference()1, a relative path to the socket 
can be given in the name if the socket resides somewhere in the subtree rooted at the container. For example name "folder/mySocket"
refers to a socket named "mySocket" inside container named "folder" in the current folder.
The methods \verb1removeSocket()1 and \verb1removeAllSockets()1 can be used asynchronously to delete sockets in the container. If the container itself
is deleted, all contained sockets are deleted.

\verb1SocketContainerListener1 objects can be attached to \verb1SocketContainers1 and they receive events 
if there are changes in the state of the container.
Access right roles can be accessed using interface \verb1Resource1 implemented by all \verb1SocketContainer1 objects.

\section{Persistence}

The class \verb1StorageBlock1 is used to access a single persistence server. Therefore, it has a single location in the HCA
domain tree and implements interface \verb1Localized1 for querying a current location. Clients can access
a \verb1StorageBlock1 using location independent references in the same way as sockets.

The class \verb1StorageBlock1 implements the interface \verb1Resource1, which can be used for setting the principals 
whose rights are used to access the persistence
server. Persistence servers can be configured by implementation-specific means to have different quotas for different principals. 
The control of access rights
to persistent storage is managed with socket containers and their access rights.
The amount of available resources can be queried with the method \verb1resourcesAvailable()1.
All attached \verb1StorageBlockListener1 objects are notified, if the amount of available resources changes. 

Interface \verb1StorageBlock1 is presented below:

\begin{mylisting}
\begin{verbatim}
class StorageBlock : public LocationIndependent, public Localized,
                     public Resource, public EventSource<StorageBlockListener> {
public:
   Integer resourcesAvailable();
   virtual ~StorageBlock();

private:
   StorageBlock();
   StorageBlock(const StorageBlock&);
   StorageBlock& operator=(const StorageBlock&);
};
\end{verbatim}
\end{mylisting}

\section{Access Control}

\begin{figure} [ht]
\centering
\centerline{\epsfxsize 0.8\linewidth \epsfbox{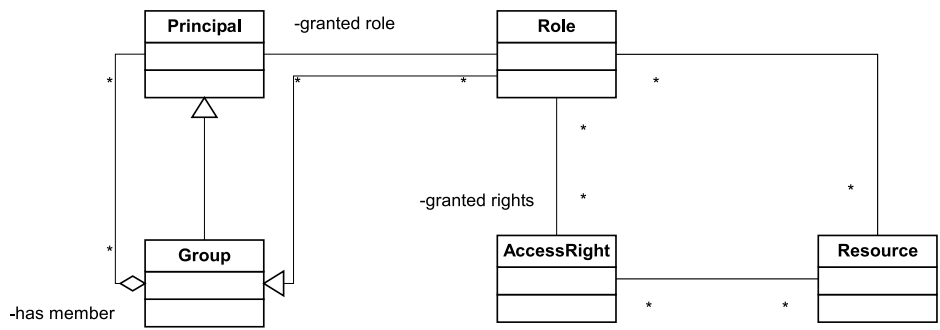}}
\caption{Access rights conceptual model}
\label{ARConcept}
\end{figure}

Figure \ref{ARConcept} presents the conceptual model of access control in HCA. The same model
is used system-wide. The access rights concepts map directly to classes in Client interface.

The concept access right in Figure \ref{ARConcept} represents any right to do something, such as,
right to write to a certain \verb1SharedVector1. Usually, it is associated with some resource it is protecting, but the possible rights
can be more abstract. Every access right is granted to a set of abstract roles, for example "owner of the socket" or
"a company employee". Roles are determined by the organization or application of the access controlled system and they form, in a sense,
viewpoints to the system.

Principals represent any entity which has an identity and can possess access rights. For example, users of the system are principals
and they can perform operations in the system. Each role is granted to a set of principals.
Principals can be grouped into larger user groups and it is possible to grant a role to a group of principals. Principals and groups
are entities independent of access controlled resources.

Access rights, roles, groups, and individual principals can be thought to form vertices of a directed graph where arcs are from
access rights to roles, roles to roles, roles to groups, groups to groups, and groups to principals. 
Every arc is interpreted as the source vertex granting
a right to the target vertex. When a principal tries to perform an operation which requires access rights, it is checked that there
exists a path in the graph from every needed access right to the principal. If this condition is met, the operation is performed, otherwise
nothing is done and user process is notified with an exception.

The access control interfaces are represented below:

\begin{mylisting}
\begin{verbatim}
class Identity : public Serializable {
   virtual vector<string> getCryptoMethods();
   virtual Key getPublicKeyForMethod(string method);
   virtual bool equals(Identity&);
   virtual ~Identity();
};


class Principal : public Identity {
public:
   virtual Key getSecretKeyForMethod(string method);
   virtual ~Principal();
};

class Group : public Socket {
public:
   void isGrantedTo(Identity&, GroupListener*);
   void getAllGrants(GroupListener*);
   void grantTo(Identity&, GroupListener*);
   void denyFrom(Identity&, GroupListener*);
   void clearRights(GroupListener*);
   void grantToAll(GroupListener*);
   virtual ~Group();

protected:
   Group();
   Group(const Group&);
   Group& operator=(const Group&);
};


class Role : public Group {
   virtual ~Role();
private:
   Role();
   Role(const Role&);
   Role& operator=(const Role&);
};

class AccessRight : public Socket {
public:
   void isGrantedTo(Role&, AccessRightListener*);
   void getGrantedRoles(AccessRightListener*);
   void grantToRole(Role&, AccessRightListener*);
   void denyFromRole(Role&, AccessRightListener*);
   void clearRights(AccessRightListener*);
   void grantToAll(AccessRightListener*);
   virtual ~AccessRight();

private:
   AccessRight();
   AccessRight(const AccessRight&);
   AccessRight& operator=(const AccessRight&);
};
\end{verbatim}
\end{mylisting}

The class \verb1Identity1 represents a global identity of any entity in the system. Basically it consists of a set of public keys. Each identity
may support multiple public key cryptographic methods and any supported method can be used to authenticate the identity. Security
interface implements an ordering of different methods according to the vulnerability of the method and more secure methods should be
preferred when available. The class \verb1Principal1 is a subclass of \verb1Identity1 and 
it encapsulates the secret key of the identity. \verb1Principal1 objects
can be used to perform operations in the system on behalf of the principal.
\fussy

   \chapter{DOMAIN TREE}
\label{ch:domaintree}

The skeleton of a HCA implementation is formed by communication nodes that form together an overlay network over the 
underlying physical network. Communication nodes are partitioned into a global domain hierarchy. Every installation
has a single root domain. 

\begin{figure} [ht]
\centering
\centerline{\epsfxsize 0.9\linewidth \epsfbox{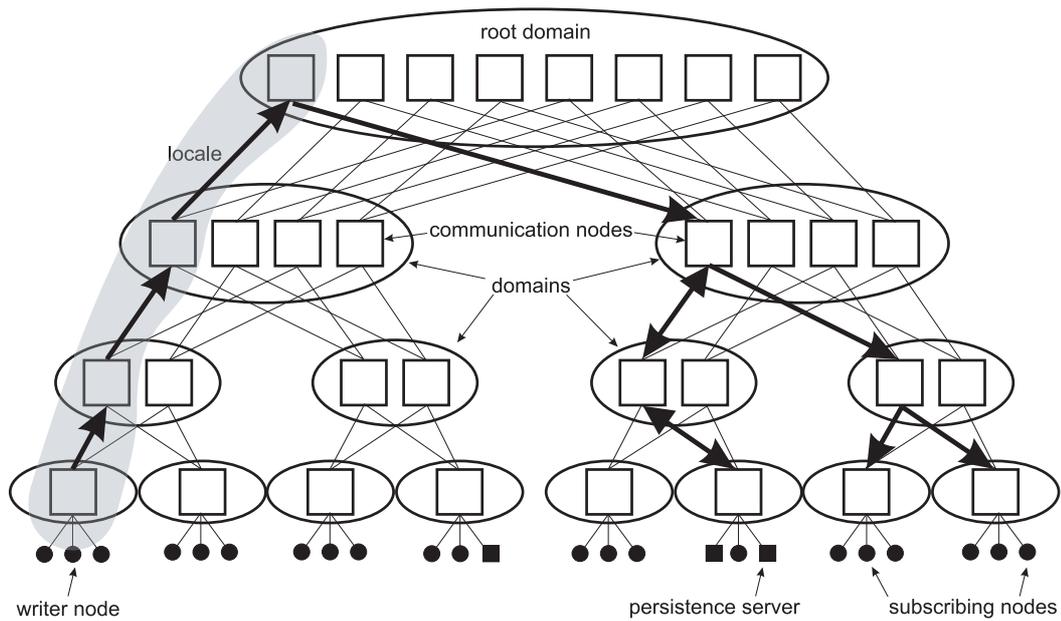}}
\caption{The HCA overlay network topology}
\label{overlaynetwork}
\end{figure}

The schematic diagram of the implementation is shown in Figure \ref{overlaynetwork}.
Client local nodes and persistence servers are attached to communication nodes residing in the leaf domains of the domain tree
as seen in the figure. We can assume that every leaf node has approximately the same number of clients and the bandwidth generated
by the clients is roughly constant. Therefore, the number of communication nodes in the leaf domains is linearly proportional to the
number of clients in the system.

It should be noted that different branches of the domain hierarchy can have different depths and arbitrary number of nodes. The diagram 
presented in Figure \ref{overlaynetwork} is completely uniform for the clarity of presentation. For the analysis of the system we can
assume without a loss of generality that the domain hierarchy is a complete binary tree as shown in the figure, each domain
having as many communication nodes as its child domains added together. Therefore, the whole overlay network has $\Theta(n*log(n))$ communication
nodes where $n$ is the number of nodes in the leaf domains combined if the tree is balanced as in the figure.

There are three requirements for an efficient structure of the domain hierarchy:

\begin{enumerate}
\item
Every domain must have a maximum of $n$ child domains where $n$ is some implementation
dependent constant. Probably good values of $n$ are somewhere between 2 and 100 depending
on the physical network topology, availability of underlying multi-cast support and the number of possible simultaneous connections 
a single node can have.
\item
If $a$ and $b$ are arbitrary child domains of domain $c$, then the number of nodes in 
domains $a$ and $b$ must be of same order of magnitude. Also the number of nodes in
subtrees starting from domains $a$ and $b$ must have the same order of magnitude of nodes.
There cannot be domains without any nodes.
\item
The number of nodes of a domain must be approximately same size as the number of
its child domain nodes added together.
\end{enumerate}

If all three condition are met by the hierarchy, the tree is balanced for a good asymptotic scalability, 
every node has a constant upper limit for the number of connections needed to other nodes and every node has on average a balanced 
load as explained later.

The above restrictions have the drawback that the domain hierarchy must be planned beforehand to avoid unnecessary maintenance work 
when the network of nodes is changed.
Also, because of the restrictions, it is necessary to create artificial domains that do not reflect any natural subset of the 
physical network, organizational hierarchy or geographical division. On one hand, this makes the politics of building 
the HCA network over many operators somewhat more difficult and, on the other hand, the mismatch between overlay network topology 
and physical network topology may prevent the full use of physical resources.

In Figure \ref{overlaynetwork}, the locale depicted contains communication nodes that reside in the same geographical area and can be
administered by the same organization. In general, vertical cross-sections of the overlay network in the figure can be interpreted
as geographical areas. Therefore, it is natural that each organization joining in the implementation of the HCA network contributes
by providing a total of $\Theta(log(n))$ nodes per user to the enclosing domains of the geographical area where the users reside. The provided
nodes high in the domain hierarchy also route information for other organizations but the situation is symmetrical
to all participating organizations and the protocol is fair to all.

The client nodes and persistence servers (small filled circles and squares in 
Figure \ref{overlaynetwork}) are spread to different geographical locations. They are connected to the leaf domain nodes each of which serve a local
area of the network. Each communication node is connected to a small number of nodes in its parent domain and each of its child domains.
The HCA inter-domain protocol is used in communication using these connections.

\section{Logical Address Space and Routing}
\label{logaddressspace}

HCA uses its own internal address space for locating entities and division of labor between communication nodes. 
Internal addresses do not
contain any location information or any other interpretation. Therefore, we call these addresses logical and they are arbitrarily assigned
to entities. 

Each address has a prefix represented by a 64-bit unsigned integer used for routing between nodes. In each domain, the address space of these
prefixes is partitioned into continuous blocks of addresses which are distributed to the nodes of the domain. Every node has a single, 
continuous block of addresses of approximately same size and the node is responsible for routing all information 
associated with addresses having a prefix in that range through the domain of the node. Each domain manages the distribution of 
address space ranges by using a domain-specific intra-domain protocol. The whole prefix address space must be mapped to domain nodes
all the time and if nodes are added or removed from the domain, the remaining nodes must handle the redistribution of the prefix address
space using the intra-domain protocol.

Each node is locally configured to find a set of nodes from its parent domain. The node can initially
communicate with some of these nodes to receive the nodes of the parent domain responsible for the same range of prefix addresses as the
node itself. Then the node forms persistent connections with the received parent domain nodes. Also, the nodes in child domains of the node
follow this protocol to connect with the node. Finally, each node is connected to a set of nodes in its parent domain and all child domains
so that the address range the node is responsible for is covered by those nodes in every neighbouring domain. 
Intra-domain protocol is used to find nodes responsible for a given address space range inside the domain. Because we have assumed that every
node is responsible for only one continuous range of addresses, the address space is divided almost equally to all nodes in the domain, 
branching degree of domains is limited by a constant and each parent domain has number of nodes only constant factor greater than
its child domains, it follows that every node needs only $\Theta(1)$ persistent connections.

For example, in leaf domains which have only a single node, the node is responsible for routing the whole logical address space but on the other
hand serves only a very specific geographical location. The root domain nodes are the complete opposite: each node handles a very specific
segment of the logical address space but is independent of the location. The intermediate domains between leafs and the root gradually
route messages from location bound nodes to more logical nodes and back.

Because communication through a socket is independent of the other sockets, it is possible to
perform all socket specific communication under a single logical address so that all communication concerning a given socket is routed
through the same virtual tree in the HCA overlay network. This tree is formed by taking the single node from each domain responsible for
the logical address bound to the socket.

\section{Socket Communication}
\label{socketcom}

In Figure \ref{overlaynetwork}, a typical case of clients communicating through a HCA socket is depicted. All messages are routed 
through the virtual tree selected by the logical address of the socket as explained in Section \ref{logaddressspace}. 
Persistent connections between nodes are drawn as thin lines in the figure and an example route followed by messages related to some socket
is drawn as thick arrows. Writers and readers of the socket can be found at the leafs of the virtual tree and, in the case of a persistent
socket, also a number of persistence servers are participating in the coordination of the socket.

Because same nodes are always used for the routing of the same sockets, caching socket data in the intermediate nodes in the virtual
communication tree can be very beneficial. Clients connect to the tree at the leafs of the tree which are closest to them.
Messages sent by a client gradually move toward larger enclosing domains until they reach the smallest common domain with the target
destination where the direction of messages turns downwards to the destination location. $O(log(n))$ hops from node to node
is needed at maximum if message has to travel to a root node before it can bounce back towards its destination.

Scalable multicast semantics follow easily
as each communication node can duplicate a message to each of its child domains interested in it with amount of work linearly
proportional only to the length of the message. Locality is also achieved automatically as messages travel only as high in the domain
hierarchy as is necessary to reach the destination. As we move from a client towards a larger domain in the virtual tree, we face nodes
with a gradually growing probability of having the needed information cached. In this kind of topology, the distance needed to travel to
the nearest cache containing the requested information is naturally inversely proportional to the popularity of the information on average.

The scope of a socket can be limited to a given domain so that messages concerning it are never routed upwards from that domain and messages
are not received from domains above the limiting domain. That domain becomes effectively a new "virtual" root domain for the socket.

\section{Identities}

In HCA, many entities like communication nodes, principals, domains, and persistence servers have global unique identities 
that can be represented as binary data. Because a global repository for managing these identities would be cumbersome to
implement, we will solve the uniqueness of the identities by using large random numbers as identities. By using large enough numbers
we can reduce the probability of an accidental name clash in normal use, for example, to the same order of magnitude 
as meteorite destroying the earth 
which renders it to a very little concern. We implement security against identity thefts by 
using authenticating authorities and public key cryptographic signing. 

\section{HCA Reference Representation}

HCA references are location independent pointers, which are used by clients to refer to sockets and persistence servers in the system.
They have a standard representation and can be serialized by any client to a binary data which can be deserialized by any client back to
a socket reference. Clients should not interpret the serialized form of the references. The serialized 
content format of a HCA reference is shown below:

\begin{center}
\scriptsize
\begin{tabular}{|c|}
\hline
Socket id : Integer\\ \hline
Contact address prefixes : list$<$Uint64$>$\\ \hline
Accepted authorities : list$<$Identity$>$\\ \hline
\end{tabular}
\end{center}

Each reference contains the identity information of the persistence server or socket pointed to by the reference. 
\emph{Socket id} is an integer value denoting the referenced entity. It must be unique among all identities authenticated by the same authorities. 
A reference contains a list of public keys of accepted
authorities that are trusted for authentication of the referenced entity. If the list of accepted authorities is empty, only
authorities trusted by the client in general are used for authentication.

A reference also contains a list of contact addresses. 
Each contact address is a 64-bit prefix address used for routing in the HCA communication tree. Multiple contact addresses are provided
for replication and they all point to the same information. Each contact address together with a socket id can be used to route to 
a \emph{socket file}, which contains 
information about the socket or storage block referenced by the reference. The socket file is published by the persistence servers
responsible for the entity referenced.

\subsubsection{Locating a Socket}

\begin{figure}[h]
\centering
\centerline{\epsfxsize 0.7\linewidth \epsfbox{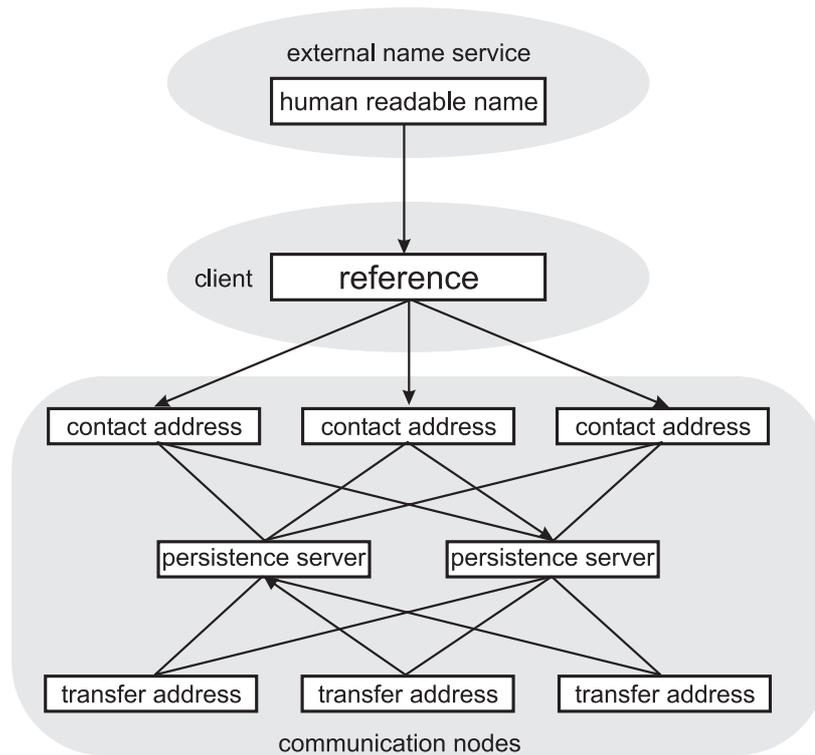}}
\caption{Locating a socket file using a socket reference}
\label{socref}
\end{figure}

The process of finding of a resource in a HCA implementation is depicted in Figure \ref{socref}. 
HCA does not incorporate human readable names for resources but an external naming service can be used for this purpose. The responsibility
of the naming service is to map textual names to location independent HCA references, which can then be used to locate the socket.

Each location independent reference contains multiple contact address prefixes which are used in finding the route through communication
nodes. Persistence servers also advertise their sockets using the same contact address prefixes used in the references pointing to
each socket. The advertisement information of each socket consists of a socket file describing the properties of the socket. Persistence
servers send these files and their updates to the responsible communication nodes, which reserve a certain amount of storage space for
socket files for each of their child domains. The storage resources are only leased for socket files and communication nodes can 
remove unused files without notice. The replication of socket files to multiple communication addresses guarantees that the socket file
can be found with high probability from at least one of the contact addresses. Each socket file also contains the current location of
each persistence server participating in the storing of the socket state. Socket files are represented as vectors just like 
\verb1SharedVector1 sockets and client interface implementations can subsribe them and receive their states and updates.
Socket files are described in more detail in section \ref{subsec:socketfiles}.

Socket files may also introduce multiple transfer addresses used for routing the payload communication of the socket in the HCA network.
This is required, because shared vectors can be very large and communication nodes on the routing path try to cache their state. When
the vector is partitioned into multiple parts and each part is routed through different communication nodes, the amount of cache per
a communication node does not limit the size of shared vectors.

   \chapter{PROTOCOLS}
\label{ch:protocols}

The protocol used by the communication nodes and persistence servers is divided into inter- and intra-domain protocols, which are
described in the following sections.

\section{Inter-Domain Protocol}

Inter-domain protocol is the communication channel between HCA clients, persistence servers, and communication nodes of different domains.
The protocol message formats and their semantics are presented in the following subsections. Message formats are defined using a simple
type system developed for this thesis. The marshalling of values of these types into a byte stream is explained below.

\subsection{Marshaling of Types}
\label{subsec:marshaling}

HCA uses big-endian style byte order for fixed-size integer types. Arbitrary size integers used in the client interface are 
marshalled in the following way: Bytes $b1,b2,..,bn$ are read until a value $bn$ that has the most significant bit set 
(unsigned byte value is greater than 127) is found. The 8th bit is then ignored and the length in bytes of the value of the integer 
is calculated $length = b_n + b_{n-1}*128 + b_{n-2}*128^2 + ..$. Then $length$ following bytes are interpreted as an integer value 
assuming the most significant byte first.
Raw data blocks and strings are marshalled in same way as arbitrary size integers. Boolean type is implemented as a byte, where a non-zero
value is interpreted as true.

Values of record types are marshalled by writing the contents of the record value consecutively. Discriminating unions are stored by
first writing an arbitrary length integer which functions as a selector and then marshalling the selected member of the union type.
List values are marshalled by first encoding the size of the list in the same way as the length of an arbitrary size integer and then 
marshalling the elements of the list consecutively to the stream. 

\verb1pair<a,b>1 denotes a generic type, which is defined to be a record containing two elements: \verb1a1 and \verb1b1.
An element of the pair type containing values \verb1a1 and \verb1b1 is presented as \verb1(a,b)1 in the notation.
We will also use other generic types, which are type valued functions, to shorten and clarify the type definitions.
Value constructors of types are written in capital letters and both type- and value-level variables in small letters. Types are written with
a capital initial letter. 
New types are defined using the notation \verb1type A = b1, where \verb1a1 is the type being defined and \verb1b1 is its definition.
For example, \verb1type maybe<a> = union[NOTHING, a]1 defines a type valued function \verb1maybe1, which takes a single type-valued 
argument \verb1a1 and produces a union type. This union type has possible values \verb1NOTHING1 or a value of type \verb1a1.

New message types are defined in a similar way: \verb1message M = def1 defines a new message format \verb1M1.
Also, \emph{has-type} relation is used in the notation: \verb1a:T1 names a variable \verb1a1 of type \verb1T1. If a variable
is introduced without a type signature, it is assumed to be type-valued.

We will use notation \verb1signed<data,key>1 to specify that the signature contains information \verb1data1 signed by
public key \verb1key1 and \verb1encrypted<data,key>1 is used when \verb1data1 is being encrypted by either public or secret key
\verb1key1. Signing and encryption of data can also be used together nested. Signing is implemented by encrypting a fixed length
message digest with a secret key of the public/secret key pair. The type \verb1authenticated<data,identity>1 is used when it is necessary
to send both an identity and the data signed with the public key of the identity.
The details of the signing and encryption depend on the cryptographic method and key used.
The following types are also used in the message format definitions. Their marshalling follows from the marshalling of the basic types.

\begin{mylisting}
\begin{verbatim}
type maybe<a>          = union [NOTHING, a]

type NetType           = String
type NetAddress        = record [NetType, RawData]

type Key               = RawData
type Method            = String
type SingleIdentity    = pair<Method, Key>
type Identity          = list<SingleIdentity>
type Principal         = list<record[Method, Key, Key]>
type Date              = record [year:Integer, month:Uint8, day:Uint8, 
                                 hour:Uint8, minute:Uint8, second:Uint8, 
                                 second1000:Uint16 ]
type Certificate       = 
   pair<authorityPublicKey:Key,
        signed<record[Integer, Identity, version:Integer, 
                      startDate:Date, endDate:Date,
                      authority:String, description:String], 
               authorityPublicKey>>

type Domain            = String
type Boundary          = String
type DomainDescription =
   record [ a:Identity, 
            signed<record[Domain, list<Boundary>], a>]
type HCALocation       = list<Domain>

type SocketType        = union  [STORAGEBLOCK, SHAREDVECTOR, MESSAGESINK, MESSAGEBUFFER, 
                                 ROLE, GROUP, CONTAINER]

type SocketRef         = record [id:Integer, contactAddrs:list<Uint64>, 
                                 authorities:list<Identity>]

type SocketFileAddr    = record [ comAddress : Uint64,
                                  socketId : Integer, 
                                  pubKey : SingleIdentity ]

type authenticated<data, id@(method, key):SingleIdentity> = pair<id, signed<data, key>>


type serverRequest<messageType> =
   authenticated<record [ SocketFileAddr,
                          requestId      : Integer,
                          message,
                          returnAddress  : maybe<SocketRef> ], 
                 clientIdentity>

type Range = union [ Integer, pair<Integer,Integer> ]
type RangeSubscription = list<record[Range, hasVersion : Integer]>
\end{verbatim}
\end{mylisting}

\label{SERVERREQUEST}
\verb1()1 is the empty type with no elements. The template \verb1serverRequest<messageType>1 is used for messages, which are sent from a
client to the persistence servers responsible for a given socket. The request is signed by the client and this client identity is used
for checking the access rights for the request on the persistence server side. The client can also provide a return address in a form
of the socket reference of a message sink socket. The persistence servers can then send the response to the requested operation to the message sink.
The \verb1serverRequest1 messages are first routed upwards in the communication node hierarchy following 
the communication address prefix in the message.
When the socket file of the socket is reached, the message is then routed downwards to all of the persistence servers listed in the socket file.

\subsection{General Message Structure}

Each message starts with a common header. The header begins with a single 32-bit unsigned integer which tells the message type. 
Communication nodes should ignore all message types they are not familiar with. This provides some future extendability of the protocol.
After the message type follows a 64 bit unsigned integer, which is a running message counter used in connection-based communication.
After the header the marshalled values of types defined in each message format follow immediately. The messages do not carry any other
form of runtime type information, but as the message type is known the values can be interpreted unambiguously.
After two communicating nodes have agreed a shared secret key for their peristent connection in the domain joining procedure, 
the communication channel is assumed to be encrypted with the shared key.

\subsection{Joining and Leaving a Domain}

\begin{figure}
\centering
\centerline{\epsfxsize 0.6\linewidth \epsfbox{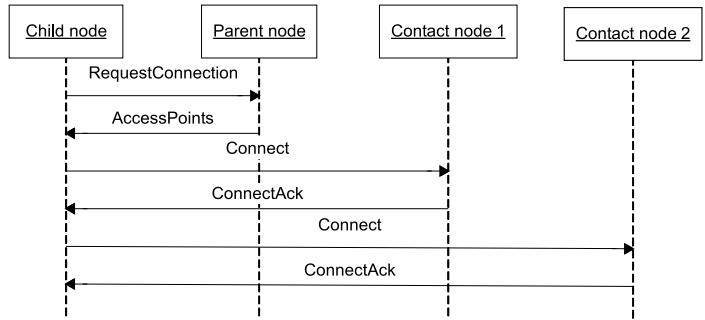}}
\caption{Sequence diagram of joining a node to its parent domain}
\label{nodejoining}
\end{figure}

Figure \ref{nodejoining} presents the sequence diagram of a communication node joining its parent domain.
First, the child node sends a message \verb1RequestConnection1 to one of the communication nodes of the parent domain and
requests the network addresses of the nodes handling the same address space as the child node. The node in the parent domain
responds with a message \verb1AccessPoints1, which contains information about the contact nodes and their replicas that the child node
needs to connect to. Then, the child node sends a \verb1Connect1 message to all contact nodes to establish a persistent connection and
contact nodes respond with \verb1ConnectAck1 messages. \verb1ConnectAck1 message contains also information about the domain
structure on top of the parent domain.

The message types for joining a domain are presented below:

\begin{mylisting}
\begin{verbatim}
message RequestConnection = 
   signed< record [ startAddress : Uint64,
                    endAddress   : Uint64 ] ,
           childNodePublicKey >

type NodeAd = record [                Identity, 
                                      NetAddress,
                       startAddress : Uint64,
                       endAddress   : Uint64 ]
type ReplicaAd = record [Identity, NetAddress]
message AccessPoints =
   signed< list<pair<NodeAd, list<ReplicaAd>>> , 
           parentNodePublicKey >

message Connect = 
   signed< record [ startAddress : Uint64,
                    endAddress   : Uint64 ],
           childNodePublicKey >

message ConnectAck = 
   signed< Encrypted <record [ sharedKey    : Key,
                               startAddress : Uint64, 
                               endAddress   : Uint64,
                               list<DomainDescription> ],
                      childNodePublicKey>, 
           parentNodePublicKey>
\end{verbatim}
\end{mylisting}

If the domain structure is changed, all nodes in domains whose paths to the root domain have changed, send a \verb1DomainChange1
message to their child nodes. If boundaries are reconfigured in some domain, the nodes propagate the change with \verb1BoundaryChange1
messages to their child nodes.

\begin{mylisting}
\begin{verbatim}
message DomainChange = list<DomainDescription>
message BoundaryChange = DomainDescription
\end{verbatim}
\end{mylisting}

The inter-domain communication between nodes is connection-based and a child node can leave its parent domain by
disconnecting.

If the address space, that the parent communication node is responsible of, changes, an \verb1AddressSpaceUpdate1 message is sent
to the relevant child nodes. Also, child nodes can notify their parent nodes with the same message type. Child nodes can also use
a \verb1RequestConnection1 message to join to new parent nodes in case of their current parents do not cover their whole address space
assignment. If the set of replication nodes changes in the parent domain, the parent node informs its child nodes by sending a new
\verb1AccessPoints1 message.

\begin{mylisting}
\begin{verbatim}
message AddressSpaceUpdate = record [ startAddress : Uint64, 
                                      endArddress  : Uint64]
\end{verbatim}
\end{mylisting}

\subsection{Generation of Contact Addresses}

Persistence servers temporarily reserve storage space from communication nodes for storing socket files. Basically, persistence servers can
create new contact addresses for socket files freely, but it is assumed that they do not produce an excessive number of replicas and choose
contact addresses randomly to achieve a good average load balancing. To prevent malicious programs wasting the resources of communication
nodes, communication nodes can use prioritization based on the current usage of the socket file, the age of the socket file, 
and the reliability of the authentication authorities of the socket to choose which files should be removed from the storage of the node
in case of shortage of resources. A communication node can also apply congestion control to its child node if it detects flooding of 
socket file creation messages. This will limit the damage caused by possible virus programs.

\subsection{Socket Files}
\label{subsec:socketfiles}

The location independence of the HCA references is achieved by using location independent contact addresses in the references.
Persistence servers advertise their sockets by publishing \emph{socket files} that are sent to the communication nodes responsible
for the address space containing the contact address of the socket. Socket files contain all essential information about the entity
they represent. The information is structured into a sparse vector to make it possible to read the contents of the file partially for efficiency
reasons. Each client interested in a socket file can subscribe parts of it by sending a \verb1SubscribeSocketFile1 message 
and receive the modifications to the subscribed parts automatically.

Each socket file vector starts with the following information:

\begin{center}
\scriptsize
\begin{tabular}{|l|l|}
\hline
0  & pubKey : Identity\\ \hline
1  & socketId : Integer\\ \hline
2  & version : Integer\\ \hline
3  & list$<$Boundary$>$\\ \hline
4  & list$<$Certificate$>$\\ \hline \hline
5  & readerRole : SocketRef\\ \hline
6  & writerRole : SocketRef\\ \hline
7  & ownerRole  : SocketRef\\ \hline
8  & lockRight  : SocketRef\\ \hline
9  & forceLockRight : SocketRef\\ \hline
10 & changeBoundariesRight : SocketRef\\ \hline
11 & destroySocketRight : SocketRef\\ \hline \hline
12 & container : SocketRef\\ \hline
13 & minReplicas : Uint32\\ \hline
14 & maxReplicas : Uint32\\ \hline
15 & SocketType\\ \hline
16 & pServers : list$<$HCALocation$>$\\ \hline
17 & lockState : union [LOCKED, UNLOCKED] \\ \hline
18-999 & RESERVED\\ \hline
1000- & Socket type specific (see below)\\ \hline
\end{tabular}
\end{center}

Indices from 0 to 17 contain general socket information common to all socket types. Indices from 18 to 999 are reserved for
future extensions and socket type specific information can be found starting from index 1000. 
Communication nodes always store data elements at indices from 0 to 17 locally,
but they have to subscribe other indices from child nodes, which are on the route leading to some of the persistence servers responsible for storing 
the socket file. It is also expected
that the communication nodes try to cache a brief version history of the previous states of the socket files.

Storage block socket files have the following additional fields:

\begin{center}
\scriptsize
\begin{tabular}{|l|l|}
\hline
1000 & availableResources : Integer\\ \hline
\end{tabular}
\end{center}

Container socket files have the following additional fields:

\begin{center}
\scriptsize
\begin{tabular}{|l|l|}
\hline
1000 & resourcesUsed : Integer\\ \hline
1001 & resourcesAvailable : Integer\\ \hline
1002 & storageBlocks : list$<$SocketRef$>$\\ \hline
1003 & numberOfContainedSockets : Integer\\ \hline
1004- & pair$<$name : String, SocketRef$>$\\ \hline
\end{tabular}
\end{center}

Message buffer socket files have the following additional fields:

\begin{center}
\scriptsize
\begin{tabular}{|l|l|}
\hline
1000 & messageCount  : Integer\\ \hline
1001 & resourcesUsed : Integer\\ \hline
\end{tabular}
\end{center}

Shared vector socket files have the following additional fields:

\begin{center}
\scriptsize
\begin{tabular}{|l|l|}
\hline
1000 & size  : Integer\\ \hline
1001 & state : Integer\\ \hline
1002- & record[transferAddress : Uint64, startIndex : Integer, endIndex : Integer, storageBlockIndices : list$<$Integer$>$]\\ \hline
\end{tabular}
\end{center}

Access right, role, and group socket types are implemented as shared vector sockets.
The element at the index 0 of the shared vector is of type \verb1union[GRANTEDALL, LIST, NONE]1, which determines whether everyone is granted
the access right,  none is granted, or the following access list is used to determine if an identity has the access right represented by the socket.
The element at the index 1 contains the number $n$ of groups that are granted the access right.
The elements at indices from 2 to $n+1$ contain each a \verb1SocketRef1 pointing to a group socket. 
Indices from $n+2$ to the end of the shared vector contain \verb1Identity1 values for all granted clients.
The elements are sorted according to the contained identities to enable faster search of a given identity.
Each element of the vector is also authenticated by the identity of the access right socket.

Message sink socket files have the following additional fields:

\begin{center}
\scriptsize
\begin{tabular}{|l|l|}
\hline
1000 & isReceiving : Boolean\\ \hline
\end{tabular}
\end{center}

A new socket file is created with a \verb1NewSocketFile1 message, see below. Communication servers will route the message to the first node
responsible for \verb1prefix1 address, which has at least one of the boundaries of the socket. If no boundaries are found, the
socket file is created in the root domain and is visible globally. If boundaries are changed, the location of the socket file is
changed accordingly. A socket file can be updated by sending a \verb1SocketFileUpdate1 message, which contains a list of modifications
to the elements of the socket file. 

\begin{mylisting}
\begin{verbatim}
type SocketData = 
   record [pubKey                : Identity,
           socketId              : Integer,
           version               : Integer,
           boundaries            : list<Boundary>,
           certificates          : list<Certificate>,
           readerRole            : SocketRef,
           writerRole            : SocketRef,
           ownerRole             : SocketRef,
           lockRight             : SocketRef,
           forceLockRight        : SocketRef,
           changeBoundariesRight : SocketRef,
           destroySocketRight    : SocketRef,
           container             : SocketRef,
           minReplicas           : Uint32,
           maxReplicas           : Uint32,
           type                  : SocketType,
           persistenceServers    : list<HCALocation>,
           lockState             : union [LOCKED, UNLOCKED]]

message NewSocketFile =
  record [prefixAddr : Uint64, 
          authenticated <SocketData, socketId>]

message SocketFileUpdate =
   signed< record [                   SocketFileAddr,
                    fromVersion     : Integer,
                    toVersion       : Integer,
                    changedElements : list<pair<index:Integer, RawData>> ],
           socketPubKey >         
\end{verbatim}
\end{mylisting}

Every modification of the socket file increases the version number of the file by one. If some client tries to attack communication nodes
by creating numerous socket files in a short time, a communication node can ask the child node, which is the source of the flood of messages, 
to limit its bandwidth by sending a \verb1SlowDownSocketFileCreation1 message. Clients and communication nodes can subscribe a subset of
the socket file by sending an \verb1AddtoSubscription1 message to the node providing the socket file. An \verb1AddtoSubscription1 message
also contains information about the old cached state of the socket file, which can be used to improve the performance of the communication
by sending only the modifications to the cached state.

\begin{mylisting}
\begin{verbatim}
message SlowdownSocketFileCreation = ()

message AddtoSubscription = 
   record [SocketFileAddr, 
           RangeSubscription] 
   
message RemoveFromSubscription = 
   record [SocketFileAddr,
           list<Range>]
\end{verbatim}
\end{mylisting}

Persistence servers can use the \verb1CheckSocketFile1 message to poll the states of the individual socket files. Communication node responsible
for the socket file will respond with a \verb1CheckSocketFileAck1 message. This is necessary,
because communication nodes are allowed to remove socket files if they are low on resources. Communication nodes can also signal
persistence servers associated with a socket of its removal by sending a \verb1SocketFileDeleted1 message to the persistence servers. 
Persistence servers can also remove a socket file by sending a \verb1DeleteSocketFile1 message to the node containing the socket file.
The messages \verb1SetBoundaries1 and \verb1Lock1 are server request messages (see explanation in Subsection \ref{SERVERREQUEST}), which are used
to set the boundaries for the socket and to lock the socket, respectively. Persistence servers respond with a single \verb1AccessRightResponse1
message to the return address specified in the server requests.

\begin{mylisting}
\begin{verbatim}
message CheckSocketFile = 
  record [ requesId     : Integer,
           sourceAddr   : HCALocation,
                          SocketFileAddr ]

message CheckSocketFileAck = 
   record [ requestId   : Integer, 
            success     : Boolean ]

message DeleteSocketFile = 
   signed< record [SocketFileAddr],
           socketPubKey>

message SocketFileDeleted = record [ socketId : Integer,
                                     pubKey   : Key ]

type LockOp  = union [ FORCE, TRY, waitTime:Integer, RELEASE]
message Lock = serverRequest<LockOp>

message SetBoundaries = 
   serverRequest<record [ removeAll  : Boolean,
                          removeList : list<Boundary>,
                          addList    : list<Boundary> ] >

message AccessRightResponse = 
   record [ requestId : Integer, 
            union [SUCCESS, ACCESSVIOLATION] ]
\end{verbatim}
\end{mylisting}

\subsection{Socket Containers and Persistence Server Protocol}

HCA communication nodes are not used to store persistent information in the system. Special clients called persistence servers attach
to the system at the edge of the HCA network and provide all persistent storage used by the sockets. Persistence servers are relatively
simple storage blocks that work together to form a large, replicated, access-controlled memory. It is possible to remove and add 
persistence servers on the fly to the system. Persistence servers store and manage
persistent sockets and implement their access control.

The following messages map directly to operations in HCA Client interface. Client sends a \verb1NewRootContainer1 message, when it wants
to create a new root container. Communication nodes route the message to the persistence servers whose addresses are enclosed in the message.
Each server responds with a \verb1NewRootContainerAck1 message whether the operation was succesful. New sockets are created with
\verb1CreateSocket1 messages and peristence servers respond with \verb1CreateSocketAck1 message accordingly.

\begin{mylisting}
\begin{verbatim}
message NewRootContainer = 
   authenticated < record [ name          : String, 
                            storageBlocks : list<HCALocation>,
                            minReplicas   : Uint32,
                            maxReplicas   : Uint32,  
                            boundaries    : list<Boundary>,
                            responseAddr  : SocketRef ],
                   clientIdentity >

message NewRootContainerAck = 
   authenticated < newContainer : maybe<SocketRef>, 
                   persistenceServerIdentity >

message CreateSocket = 
   serverRequest < record [ name         : String, 
                            quota        : Integer,
                            initialOwner : Identity, 
                            SocketType ] >

message CreateSocketAck =
   authenticated < record [ requestId : Integer, 
                            newSocket : maybe<SocketRef> ], 
                   containerIdentity >
\end{verbatim}
\end{mylisting}

There are also several message types for operations that clients can perform on sockets. \verb1RemoveSocket1 can be used to remove a socket.
\verb1SetMinReplicas1 and \verb1SetMaxReplicas1 are used to change the replication settings. \verb1AddStorageBlock1 and \verb1removeStorageBlock1 
change the set of peristence servers associated with a socket. \verb1ChangeStorageBlockUser1 changes credentials that are used for accessing
a given persistence server storing the socket. \verb1SetQuota1 changes the quota setting for the socket.
All of these operations are defined as a \verb1serverRequest1 template, which was introduced in section \ref{subsec:marshaling}.
Persistence servers respond to each of these messages with a \verb1ContainerOperationResponse1 message to the message sink return address
specified in the server request.

\begin{mylisting}
\begin{verbatim}
message RemoveSocket = 
   serverRequest < union [REMOVEALL, name : String] >

message SetMinReplicas = 
   serverRequest < Uint32 >

message SetMaxReplicas = 
   serverRequest < Uint32 >

message AddStorageBlock = 
   serverRequest < record [ storageBlock : SocketRef, 
                            UserId : Identity ] >

message RemoveStorageBlock = 
   serverRequest < record [ storageBlock : SocketRef, 
                            UserId : Identity ] >

message ChangeStorageBlockUser = 
   serverRequest < record [ storageBlock : SocketRef, 
                            UserId : Identity ] >

message SetQuota = 
   serverRequest < record [ name : String, quota : Integer ] > 

message ContainerOperationResponse = 
   record [ requestId : Integer, 
            union [SUCCESS, ACCESSVIOLATION] ]
\end{verbatim}
\end{mylisting}

\subsubsection{Replication and Persistence Server Coordination}

We have two distinct goals for replication: To provide better availability of resources in case of partial failure and 
to guarantee that information is never lost by accident. It should be noted that the replication doesn't protect the
data from intentional attack if the identity of the owner of the data is stolen.

\begin{mylisting}
\begin{verbatim}

\end{verbatim}
\end{mylisting}

\subsection{SharedVector Socket Communication}

All message formats used in shared vector socket communication are shown below. Clients and communication nodes in the route of a shared vector
communication use a \verb1ChangeSubscription1 message
to change the set of indices for which they want to receive the state updates. These messages contain also a version number of a cached
state that the client or a communication node may possess. This way the sender can make a decision whether it is easier to send all state updates
succeeding the cached version or only the latest state in whole. The actual shared vector updates are contained in \verb1Update1 messages send
by the writer of the vector. Persistence servers respond to updates with \verb1Commit1 messages. A \verb1Snapshot1 message can be used to
request the most recent state of a shared vector without subscribing it. Communication nodes may respond to subscriptions with
a \verb1SubscriptionError1 message if the operation was not succesful.

\begin{mylisting}
\begin{verbatim}
message ChangeSubscription = 
   record [SocketFileAddr, 
           add    : union [ALL, RangeSubscription],
           remove : union [ALL, list<Range>] ]

message Update = 
   signed< record [ SocketFileAddr,
                    transferAddr : Uint64,
                    newState : Integer, 
                    list<pair<index:Integer, RawData>> ],
           socketPubKey >

message Commit = 
   record [ state         : Integer, 
            storageServer : SocketRef]

message Snapshot = SocketFileAddr

message SubscriptionError = 
   record [socketId  : Integer,
           socketKey : Key ]
\end{verbatim}
\end{mylisting}

\subsection{MessageSink Socket Communication}

All message formats used in message sink socket communication are shown below. A new message to a message sink socket is sent
with a \verb1Message1 message. It cointains the address of the target message sink and the address of message buffer used to store
the message in case of a failure. A fallback message sink address is also provided and if the primary destination cannot be contacted,
the message is eventually routed to the fallback address. \verb1maxTimeInMs1 is the maximum time used for storing the message if the destination
cannot be reached immediately. The owner of the message sink can use a \verb1SetMaximumMessageLength1 message to change the maximum 
length parameter for accepted messages. The client can also signal with \verb1StartReceiving1 and \verb1StopReceiving1 messages that it
wants to start or stop receiving the contents of the message sink.
A \verb1ConsumeMessage1 message is used to signal the sender that the message has been read and can be removed from the message buffer and/or
caches of the communication nodes.

\begin{mylisting}
\begin{verbatim}
message Message = 
   authenticated< record [ SocketFileAddr,
                           messageData     : RawData, 
                           messageBuffer   : SocketRef, 
                           fallbackAddress : SocketRef, 
                           maxTimeInMs     : Integer],
                  clientIdentity >

message SetMaximumMessageLength = 
   serverRequest< messageLength : Integer >

message StartReceiving =
   authenticated <SocketFileAddr, clientIdentity>

message StopReceiving = 
   authenticated <SocketFileAddr, clientIdentity>

message ConsumeMessage = 
   serverRequest<()>
\end{verbatim}
\end{mylisting}

A \verb1ClearMessage1 message is sent to a message buffer socket to request a removal of a given message or all messages.
The message buffer responds with a \verb1MessageBufferResponse1 message.

\begin{mylisting}
\begin{verbatim}
message ClearMessage = 
   serverRequest<union [ALL, index : Integer]>

message MessageBufferResponse = 
   record [ requestId : Integer, 
            union [SUCCESS, ACCESSVIOLATION] ]
\end{verbatim}
\end{mylisting}

\subsection{Access Right Protocol}

Group, role, and access right sockets are all used in the same way. 
A new identity is granted the access right or a group/role membership, represented by the 
socket, by sending a \verb1grantTo1 message to the persistence servers containing the socket. A \verb1DenyFrom1 message is used to 
remove an identity from the set of identities that are granted the access right or a group/role membership. 
A \verb1ClearRights1 message is used for removing
all identities that are granted the access right and a \verb1GrantToAll1 message is used to grant the access right to everyone. 
The persistence servers will
answer each request by sending a single \verb1AccessRightResponse1 message to the message sink socket provided in the \verb1serverRequest1.

\begin{mylisting}
\begin{verbatim}
message GrantTo = ServerRequest<grantTo : Identity>

message DenyFrom = ServerRequest<DenyFrom : Identity>

message ClearRights = ServerRequest<()>

message GrantToAll = ServerRequest<()>

message AccessRightResponse = 
   record [ requestId : Integer, 
            union [SUCCESS, ACCESSVIOLATION] ]
\end{verbatim}
\end{mylisting}

\subsection{Node Failure Recovery}

The persistent connections between neighbouring nodes use keep-alive messages to detect failures in communication and other nodes. If node detects
that it is disconnected from its parent node, it will choose the first working replica of the parent node and send an \verb1ActivateReplica1
message to it, which signals the parent node replica to join its own parent domain with the communication address prefix range of the child node.
The replicas of the parent node are always sorted according to the priority and child node should always choose a working replica with highest
priority. The Priorities are set in the parent node configuration information.
Nodes should use random trigger value for the accepted delay of the keep-alive messages so that in case of a large network failure a burst 
of reconnection messages is avoided. 
When the child node is able to re-establish a connection to a parent node with higher priority than the current replica in use, 
it will gradually move all socket subscriptions back to the node with higher priority and eventually to the primary server when it becomes available.
The child node can also use  an \verb1ActivateReplica1 message to signal the parent node to stop 
sending other socket related messages except the active subscriptions.
If the set of replicas for parent node changes, it will notify its child nodes with a \verb1ReplicaUpdate1 message, which contains a list of replicas
with the primary node at index 0 and all other replicas following in priority order.

\begin{mylisting}
\begin{verbatim}
message ActivateReplica = union [ACTIVATE, DEACTIVATE]

message ReplicaUpdate = list<ReplicaAd>
\end{verbatim}
\end{mylisting}

\section{Intra-Domain Protocol}

Intra-domain protocol is a domain-specific protocol and architecture for managing the nodes inside a domain. Domains in the leafs
of the domain tree are localized but the root domain and other domains high in the hierarchy are large and can span multiple
organizations. Small domain nodes can be managed with a local configuration of each node but larger domains need more complex
solutions. Intra-domain protocol is responsible for implementing at least the following functions:

\begin{itemize}
\item
Joining a new node to the domain and transferring the responsibility of handling a range of logical prefix address space from some 
old nodes to the new node. Each node should handle a range of addresses of approximately the same size.
\item
Leaving a node from the domain and transferring the responsibility of handling the range of logical prefix address space assigned
to the node to some other nodes. 
\item
Provide each node of the domain with a capability to query which nodes are responsible for handling a given range of logical
prefix address space.
\item
Configure nodes with boundaries associated with the domain.
\end{itemize}

For large domains, some DHT system could be used for a basis of an implementation of the intra-domain protocol. For example, Chord provides
the basic operations of partitioning responsibilities to a set of nodes and finding the correct node. Intra-domain protocol must also 
incorporate some security mechanism for preventing malicious nodes entering the HCA network. A detailed explanation of some intra-domain
protocol is outside the scope of this thesis.

\section{Protocol mappings}

Inter-domain protocol was specified in an abstract manner by defining packet formats and describing functionality of the communication nodes. 
This was done mainly because the low-level details of the protocol implementation 
are outside the scope of the thesis and some modularity is gained by separating the abstract functionality of the communication nodes from
specific network technology details. 
It can even be thought that several mappings could exist simultaneously to the same underlying 
technology which could possess different sets of features. It is possible to span a single HCA overlay network over multiple network 
technologies, for example, Internet and some mobile wireless network technology.

The protocol mapping should at least support fragmenting messages into multiple packets and multiplexing messages to the
same channel to provide a continuous and smooth operation and a good responsiveness for small messages among large ones.

Because persistent connections of HCA nodes have a long average lifespan and there is a lot of redundant information in consecutive 
packets, it is beneficial to use some sort of online algorithm for compressing the contents of packets before encrypting 
and signing them. Knowledge of the abstract protocol can help the compression significantly, because many messages between nodes
contain the same socket identities and other message fragments repeated over and over again. This kind of repetitive information can be 
replaced with shorter tags agreed on
by both ends of the connection. The protocol mapping may add a meta-protocol for managing this kind of abbreviations. Of course, general
compression methods could be used such as dynamic Huffman codes \cite{DYNHUFF}.

\subsubsection{Internet mapping guidelines}

It is clear that mapping the HCA protocol on top of TCP/IP network will be the most important case. 
The most natural implementation will be to map persistent connections to TCP connections and light-weight packets to UDP datagrams and
IP multicast.

It is also conceivable to use different strength cryptographic algorithms for intranet HCA domains and public Internet 
connections. In some cases, the physical connection itself can be impossible to intercept which makes software  
protection unnecessary.

\section{Communication Node}

\begin{figure} [ht]
\centering
\centerline{\epsfxsize 0.4\linewidth \epsfbox{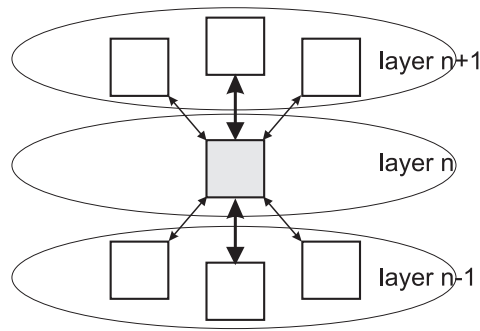}}
\caption{Single communication node persistent connections}
\label{singleNode}
\end{figure}

A single communication node and its persistent connections to neighbouring nodes are presented in Figure \ref{singleNode}.
Every node is configured with a network address for contacting some node in its parent domain. The node asks the parent node
to provide addresses for those nodes in the parent domain which are responsible for the same address space as the node. Then
the node forms a persistent connection with these nodes. Also, nodes from each child domain of the node contact the node in the same way.

\subsection{Node configuration}

Communication nodes in HCA need to be locally configured before attaching them to the overlay network. A large portion
of node configuration is implemented locally and this makes the setup of HCA network somewhat less flexible. This is 
necessary to protect the system from attacks that can globally affect the network. When individual nodes are 
compromised, it should not be possible to break the overlay network globally, because the configuration is divided into
many local parts. The reduced flexibility of setting up the network is not crucial, because majority of the nodes are
expected to be under adminstration of some service provider and HCA nodes are not designed to be added and removed
constantly like in more light-weight peer-to-peer systems which use peers located in volatile home computers.

Every node has at least the following configuration parameters:

\begin{itemize}
\item
Information needed to join the intra-domain system.
\item
A domain identity needed for joining as a child node to the parent domain.
\item 
A node identity.
\item
Domain name as a human readable string.
\item
A list of cryptographic methods accepted for node authentication.
\item
A list of parent domain node identities and their network addresses for joining to the parent domain.
\item
A list of identities of all accepted child domains and the minimum cryptographic method required for protecting
the communication with each of them. For example, in protected intranets it could be possible to optimize communication
by using a weaker encryption method. Each child domain is also given a relative priority to other child domains. The priority
should be proportional to the number of nodes in the child domain.
\item
A flag indicating if underlying network multicast can be used to optimize the duplication of messages to the child domains.
\end{itemize}

HCA provides Context interface for nodes to acquire the configuration information in a system independent way.

The interface for setting the configuration information is implementation-dependent and outside the scope of this
specification. It is possible to define a standard file format for configuration information. This will
ease the configuration of multiple nodes because a uniform method is applied across the nodes and settings
can be easily copied from other nodes.
However, the configuration information of a node contains its secret key and the necessary information to steal the identity
of the node. Therefore, it is necessary to keep the access to this information limited.

\subsection{Node Cache}

Communication nodes should use local space to cache SharedVector states and updates, so that they can quickly serve
new subscriptions from child nodes. Also socket files, socket file server references, and access right information 
should be cached. Nodes can
also save usage statistics of sockets to further enhance the use of caching resources. 

The actual caching algorithm is dependent on the node implementation, but it should maximize some kind of utility function
based on usage patterns of data. Data with high usage and low update rate is particularly suitable for
caching. Another basic principle is that local resources should be in maximum possible use,
which means that cache pruning should be done only when local resources are about to end. Also, terminating the 
subscription of data does not necessarily mean removal of the data from the node cache, because protocol enables the use
of cached states when re-subscribing to the socket. If the sending node has cached enough older states to recover the
cached state of the receiving node to the current state with less communication than by sending the most recent state in its entirety,
the cached state can be put to good use.

Nodes should also keep data subscriptions alive to the last possible moment when there are not enough
bandwidth to receive all updates anymore. This means that nodes should not terminate subscriptions automatically even when
there are no child nodes subscribing the socket. This caching of subscriptions should also be prioritized by maximizing some
implementation dependent utility function. For example, it is most beneficial to keep popular but low bandwidth 
subscriptions alive.

\section{Shortcut Connection Protocol}

\begin{figure}[h]
\centering
\centerline{\epsfxsize 0.7\linewidth \epsfbox{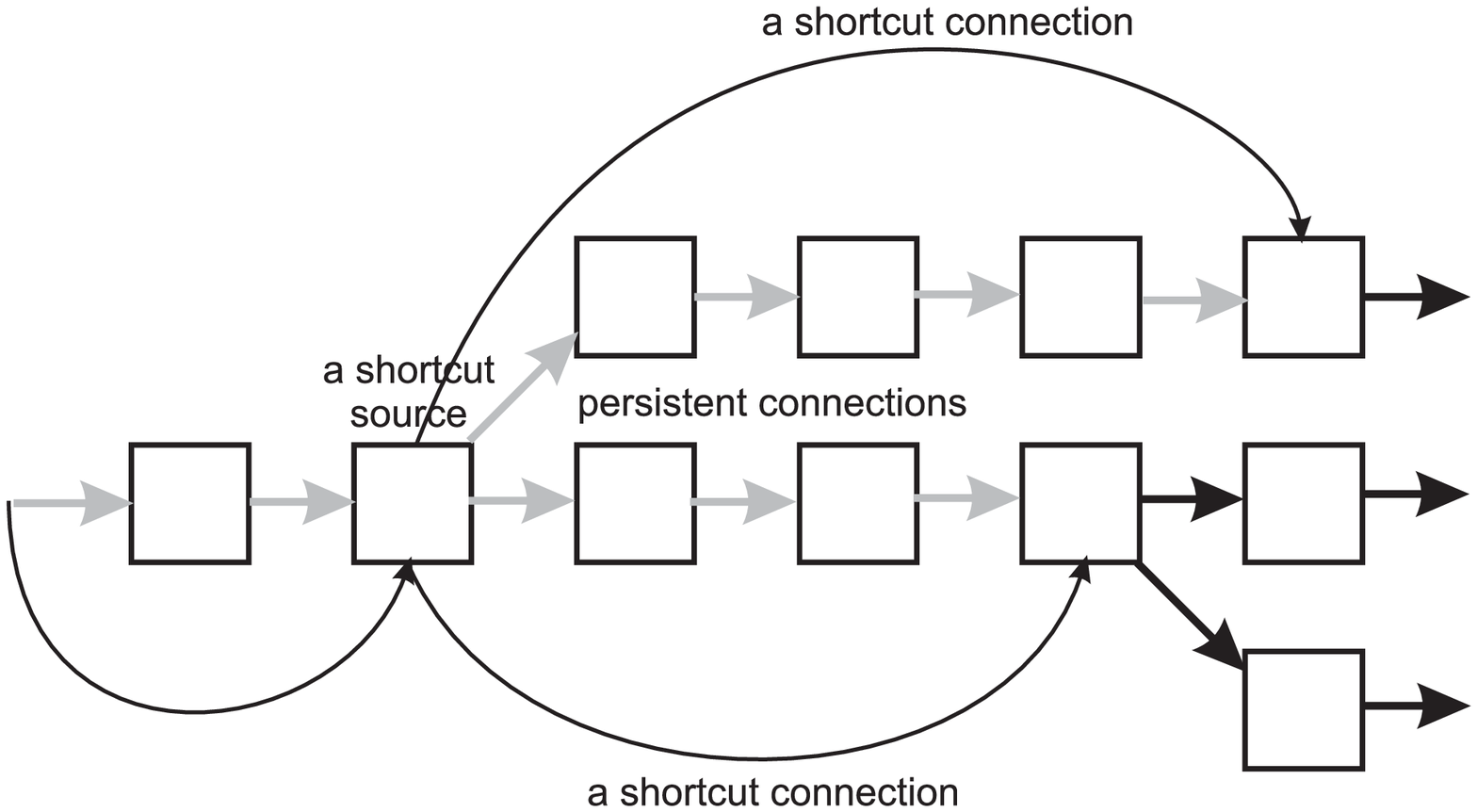}}
\caption{Using shortcut connections to minimize latency}
\label{shortcut}
\end{figure}

HCA network messages related to \verb1SharedVector1 and \verb1MessageSink1 sockets can travel through many nodes even if there are only
few readers of the \verb1SharedVector1 or few writers of the \verb1MessageSink1 socket. This increases the latency of messages and the total
load of nodes unnecessarily. HCA inter-domain protocol incorporates so-called \emph{shortcut connections} to avoid this problem.
This enables nodes, which are routing certain messages from a single source to a single destination, to break out from the pathway of
the messages and establish a more direct routing of messages. The new temporary connections are called shortcut connections. Chained shortcut
connections can also be combined to form longer shortcuts that skip multiple nodes at a time. To simplify the algorithms needed to ensure
the consistency between many concurrent nodes, the handling of the logic of each chain of shortcut connections is done by the first node
in the chain. This node is called the \emph{source node} and the last node in the chain is called a \emph{target node}.
Each shortcut is always associated with a socket and only messages related to the socket are routed through the shortcut. If multiple shortcuts
connect two nodes, it is possible for the nodes to share the same network connection for both shortcuts.
Shortcut connections add overhead in form
of number of connections needed between nodes and it is possible for nodes to reject the creation of a shortcut if they cannot handle more
connections at a time. Sometimes a node can function as a firewall or connect two incompatible network technologies, in which case it
will not initiate a shortcut connection.
A graphical depiction of shortcut connections is shown in Figure \ref{shortcut}.

\begin{figure}
\centering
\centerline{\epsfxsize 0.6\linewidth \epsfbox{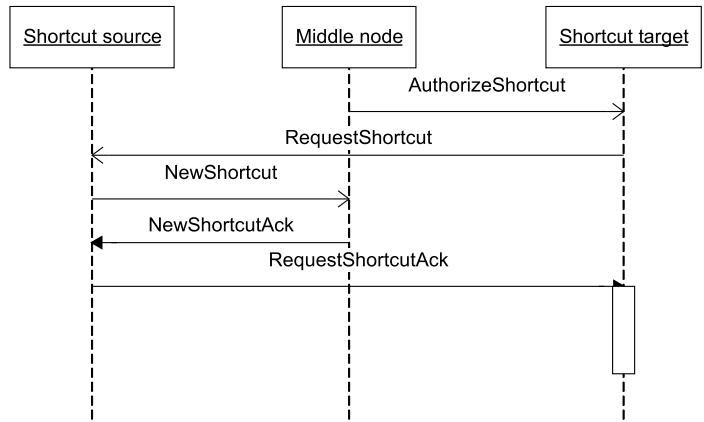}}
\caption{Creation of a new shortcut connection}
\label{newShortcut}
\end{figure}

A sequence diagram of the creation of a new shortcut connection is shown in Figure \ref{newShortcut}. At first, a node requests
to be left out from the normal path of the messages of a socket. It sends a \verb1AuthorizeShortcut1 message to the target node, which
then decides if it accepts the request for creation of a shortcut connection. If it accepts, it then sends a \verb1RequestShortcut1 message
to the source node along with the credentials received from the middle node. If the source node accepts the new shortcut connection,
it will notify the middlenode with \verb1NewShortcut1 message and after receiving \verb1NewShortcutAck1 response, it establishes the new
connection by sending \verb1RequestShortcutAck1 message to the target node. Chained shortcuts are created with the same protocol.

The message types for creation of a shortcut connection are presented below:

\begin{mylisting}
\begin{verbatim}
type ShortcutAuthorization =
   authenticated< record [ requestId    : Integer,
                           targetPubKey : Identity,
                           socketId     : Integer, 
                           socketPubKey : SingleIdentity,
                           source       : NetAddress ],
                  middleNodeIdentity>

message AuthorizeShortcut = ShortcutAuthorization

message RequestShortcut =
   authenticated< record[ShortcutAuthorization, target : NetAddress],
                  targetNodeIdentity >

message NewShortcut = requestId : Integer

message NewShortcutAck = requestId : Integer

message RequestShortcutAck = 
   authenticated <encrypted <requestId : Integer, sharedKey : Key, targetNodeIdentity>,
                  sourceNodeIdentity>
\end{verbatim}
\end{mylisting}

\begin{figure}
\centering
\centerline{\epsfxsize 0.6\linewidth \epsfbox{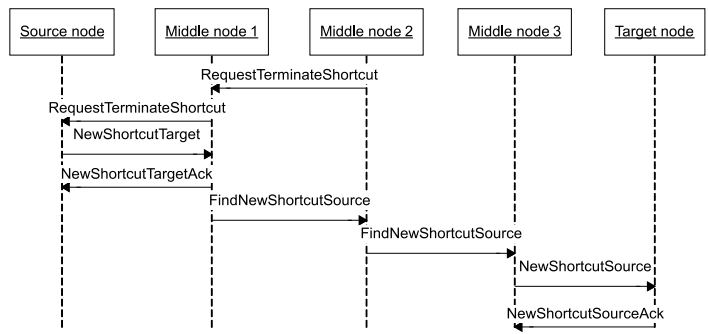}}
\caption{Termination of a shortcut connection by a middle node}
\label{terminateShortcutMiddle}
\end{figure}

A shortcut connection can be terminated by one of the skipped middle nodes, the target node, or the source node.
Middle nodes may request termination because they require the information communicated through the shorcut connection, for example, in case of
a new subscription. The target node may terminate the shortcut if it terminates the subscription of the socket communicated through the shortcut.
Both the source and the target nodes can also terminate the shortcut if they run low on resources. If multiple nodes want to terminate the shortcut
concurrently, the source node functions as a control point for choosing which request is served first. In case of failures, the nodes can always
revert to using the normal communication path.
A sequence diagram of the termination of a shortcut connection initiated by a middle node is shown in Figure \ref{terminateShortcutMiddle}.

\begin{figure}
\centering
\centerline{\epsfxsize 0.6\linewidth \epsfbox{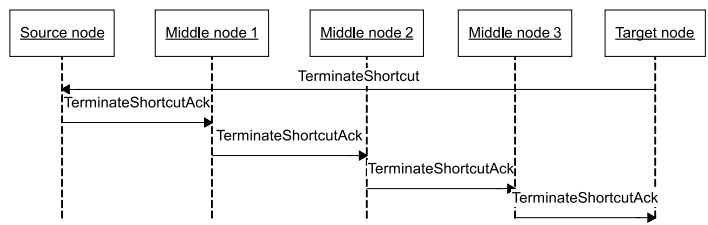}}
\caption{Termination of a shortcut connection by the target node}
\label{terminateShortcutTarget}
\end{figure}

A sequence diagram of the termination of a shortcut connection initiated by the target node is shown in Figure \ref{terminateShortcutTarget}.

\begin{figure}
\centering
\centerline{\epsfxsize 0.6\linewidth \epsfbox{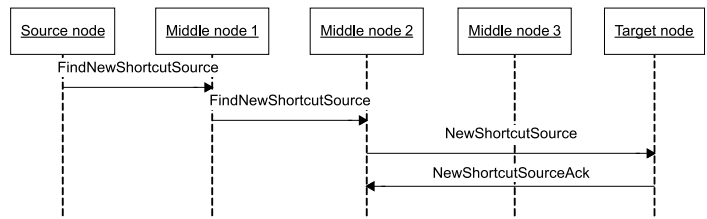}}
\caption{Termination of a shortcut connection by the source node}
\label{terminateShortcutSource}
\end{figure}

A sequence diagram of the termination of a shortcut connection initiated by the source node is shown in Figure \ref{terminateShortcutSource}.

The message types for termination of shortcut connections are presented below:

\begin{mylisting}
\begin{verbatim}
message RequestTerminateShortcut = 
   record [requestId    : Integer, 
           socketId     : Integer,
           socketPubKey : SingleIdentity, 
           middleNode   : Identity, 
           newTarget    : maybe<pair<newTarget:Identity, NetAddress>> ]

message NewShortcutTarget = 
   authenticated < record [requestId : Integer, NetAddress], 
                   sourceNodeIdentity >

message NewShortcutTargetAck = 
   authenticated < record[requestId : Integer, sharedKey : Key], 
                   newTargetNodeIdentity >

message FindNewShortcutSource = 
   record [ allowed      : Boolean, 
            requestId    : Integer, 
            socketPubKey : SingleIdentity,
            socketId     : Integer ]

message NewShortcutSource = 
   authenticated < record [ requestId    : Integer, 
                            socketPubKey : SingleIdentity, 
                            socketId     : Integer,
                            NetAddress ], 
                   newSourceNodeIdentity >

message NewShortcutSourceAck =
   authenticated < encrypted< record [sharedKey:Key, NetAddress], 
                              newSourceNodeIdentity>, 
                   targetNodeIdentity >

message TerminateShortcut = 
   record [ requestId    : Integer, 
            socketId     : Integer,
            socketPubKey : SingleIdentity ]

message TerminateShortcutAck = 
   record [ requestId    : Integer, 
            socketId     : Integer,
            socketPubKey : SingleIdentity ]
\end{verbatim}
\end{mylisting}

   \chapter{DISCUSSION}
\label{ch:discussion}

In the following sections the specified system is analyzed and evaluated against the criteria set in Chapter \ref{ch:problemDef}.

\section{Analysis of Scalability}

As we already stated in Chapter \ref{ch:domaintree}, in a balanced binary tree implementation of the domain hierarchy where each
layer of domains has $\Theta(n)$ communication nodes where $n$ is the number of the leaf nodes combined, each node needs only
a constant number of connections to other nodes. To show that the system is indeed scalable, we need to show that each node needs
to perform a constant amount of work on average to provide a constant bandwidth to each user.

In Chapter \ref{ch:domaintree} it was required that every user of the HCA implementation must contribute $\Theta(log(n))$ resources in the form
of communication nodes
to the implementation. This can be interpreted as a reserved constant bandwidth through $O(log(n))$ layers of enclosing domains.

If we consider only communication associated with a single socket, each message associated with it travels twice through the domain
hierarchy in the worst-case: first from the source client to the root domain node and from there to the target nodes. Because
the duplication of the message to all child domains at each level of the domain tree can be done in constant time and at each
level there are a maximum of $k$ nodes mediating the message where $k$ is the number of active readers and writers of the file, this
bandwidth can be subtracted from the bandwidth reserved for the writers and readers of the socket at each level of the domain tree.
The worst-case scenario is the case where each client is communicating with a different client through the root domain. Then the whole
bandwidth generated by clients flows through each layer of the domain tree by consuming $\Theta(log(n))$ routing resources per client.
Therefore, if we can show that the load is evenly balanced to all nodes at every layer of the domain tree, the system can provide
a constant bandwidth to each client.

\subsection{Average Load of a Node}

In the absence of malicious use, each socket has a randomly chosen logical address. Therefore, we can assume, that in every domain,
communication associated with each socket is the responsibility of a random node. Let us assume that there are $n$ nodes in
the domain and $k$ sockets which have active readers or writers under the domain. We will also assume that $k>>n$ and approximately
the same bandwidth is associated with each socket. This is a very basic combinatorial problem of $n$ boxes and $k$ balls randomly
dropped into them. If we examine only one of the nodes, the probability $p$ of each socket falling into its responsibility is 
$\frac{1}{n}$ and the number of total sockets served by the node follows a binomial distribution. The mean load is $\frac{k}{n}$
sockets and standard deviation of the load is $\sqrt{\frac{k}{n}*(1-\frac{1}{n})}$. Because $k>>n$ we can use the normal approximation
of the binomial distribution and therefore 99.9\% of the cases lie within 4 standard deviations from the mean. For example, if we
have $n=100$ and $k=10000$, the load of a node is between 60 and 140 sockets with 99.9\% probability. By splitting the communication
into small enough streams compared to the capabilities of a single node, we can guarantee with high probability that the load is
evenly balanced among nodes.

It may be possible to devise a distributed denial of service attack against a single node by using multiple sockets which have the 
same prefix of a logical address. This would slow down the communication of all sockets which have the same prefix.

\subsection{Latency}
\label{subsec:latency}
As already explained in Section \ref{socketcom}, in the worst case $O(log(n))$ hops between communication nodes is needed for messages
to reach their destinations. The total latency of the messages, however, is dependent also on the geographical distance travelled by
the messages because of the finite speed of light and number of devices in the underlying network the messages need to pass through in
their way. In systems like the basic Chord, which do not exploit locality in their design, the number of hops between nodes 
needed is linearly proportional to the distance travelled. For example, if the worst-case distance between nodes in the network is denoted
by $D$, the worst-case distance travelled by a query is $O(D*log(n))$. The same holds true for the average distance $d$.

Because HCA domain structure can be designed to correspond to geographical location and the underlying network topology, we can achieve
better average and worst-case latencies than, for example, the basic Chord system. By assuming a balanced binary tree domain topology and 
an approximation of network distances where the average and worst-case distances between nodes is halved when we move one step lower
in the domain hierarchy\footnote{In the case of binary tree domain hierarchy, this follows from the model where 
nodes are distributed randomly on one-dimensional segment of a line which is split in the middle to two equal subdomains and cartesian
distance is used. The assumption also holds true for a quadtree domain structure which subdivides a square to 4 identical smaller squares
and two-dimensional cartesian distance is used between randomly distributed nodes.} we get the worst-case latency as a geometric series
$2*(D+\frac{1}{2}D+\frac{1}{4}D+\frac{1}{8}D+..)$ yielding $4D$ when the message travels through the root domain and 
at every step the distance between consequtive nodes is the maximum possible. If the average distance between 
two nodes is $\frac{1}{2}D$ then the
average distance travelled by messages through the root domain is $2D$. For local communication, the distance is linearly proportional to
the diameter of the smallest enclosing domain which contains both the source and the targer client of the message. 
When the socket does not have many subscribing clients, shortcut
connections can significantly shorten the distance travelled by messages.

\label{latopt}
It would be possible to halve the worst-case and average distances easily by sending the messages from source client to the root domain
using only nodes in the same locale as the source by using different logical address for the socket in every enclosing domain around
the source client. This design was discarded, because it would invalidate the caches of intermediate nodes every time the source
migrates to another location and shortcut connections already provide shorter latency for unpopular sockets.

\subsection{Fairness}

By fairness we mean the guarantee by the system that every user is provided a bandwidth proportional to the investment made. Fairness is 
implemented in the HCA routing by each node trying to provide its child nodes an upload bandwidth proportional to the number of nodes in
the child domain. This way each domain should get a total bandwidth which is linearly dependent on the size
of the domain. Download bandwidth from parent nodes should be balanced equally to all parents. Download and upload bandwidths should also be
balanced. If node is receiving data through a shortcut connection, the bandwidth of this is added to 
the bandwidth of the persistent connection where the data would otherwise come.

Of course, nodes should always try to exploit all resources available, which means that if some nodes do not use
their full bandwidth, the remaining surplus should be divided to other nodes. Nodes should use a reasonably short time window for
measuring and dividing the bandwidth to be able to quickly adapt to new situations.
It is also impossible for organizations to use their nodes high in the domain hierarchy to favor their own domain in the long run, because
this will be detected by other organizations.

\section{Analysis of Fault Tolerance}

HCA uses replication of communication nodes and persistence servers to provide fault tolerance in the system.
Replication was chosen instead of erasure coding \cite{ERASURE} to provide fault tolerance for persistent data because of the simplicity
of implementation and because of the assumption of relatively high availability of persistence servers. Erasure coding would
have provided better fault tolerance for the same amount of storage space used, but it 
has been argued in \cite{ERASURE} that the advantages of erasure coding compared to replication
are small or non-existent when the nodes have high availability and the system is viewed as a whole.

\subsubsection{Failure Model}

We will assume that any set of communication nodes can become unreachable simultaneously. The reason for a malfunction 
can be either network or computer fault. It is also possible for the network to split so that nodes are divided in to
two distinct networks that cannot reach the other network but keep working independently.

Let us assume that each node in the system is independently unreachable with a probability $p$ for any reason. (The independence of
failures from each other is not a realistic model but we want to keep the analysis simple.) If we assume that each component
is replicated by $k$ replicas and there are $log(n)$ nodes in the pathway of the messages in the worst case, the probability of 
a message not reaching its destination is $1-(1-p^k)^{log(n)}$.

Because nodes work independently from each other in the HCA overlay network and there are no failure bottlenecks which would affect
other nodes globally, all node failures are localized either to a certain area of the network or a small set of logical addresses
depending on the location of the failing node in the network.

\section{Results}

In this thesis we succeeded in achieving the scalability for the design of HCA: With $\Theta(n*log(n))$ symmetrical communication nodes 
where $n$ is the
number of users we can build an overlay network in which every node processes on average only $\Theta(1)$ bandwidth, uses $\Theta(1)$ storage space and
has $\Theta(1)$ persistent connections, assuming that the topology of the network is well-designed. Intra-domain protocol is not taken into account
here because it does not have an effect to the normal transfer functions of the HCA unless the topology is changed or in case of a node failure. 
The HCA network provides every user with 
an upload and a download bandwidth independent of the number of other users and a feasible business model for building the network.
The constant factor overhead of the distribution protocol should be reasonably low because basic connections between nodes are persistent
and the redundant information should be possible to compress at the protocol mapping level.

The client programming interface for using HCA has only few C++ classes and the separation of 
inter- and intra-domain protocols simplify the implementation of the system. Clients can identify services and shared data
using socket references that are location and migration transparent. The API also hides the extensive replication and caching used by
the system and masks many types of failures from the client. The abstraction level of the API should be well-balanced between harnessing
the resources provided by typical underlying networks and still providing powerful distribution primitives.

HCA uses replication with communication nodes and persistence servers to achieve 
protection from accidental loss of persistent data
and a high probability of continuing to function in case of normal network and node failures. 
The exploitation of locality reduces
fault-tolerance somewhat compared, for example, to Chord system because network failures are often local and the random graph
topology of Chord with vertice degree of $log(m)$ \cite{CHORD} where $m$ is the number of vertices has outstanding robustness properties.
The HCA inter-domain protocol also implements a recovery protocol for transparent fixing of replicated nodes.

Efficient utilization of locality and knowledge of the underlying network topology is supported by HCA. Accessing local resources is 
guaranteed to use only local communication. HCA also uses caching of shared data and it is guaranteed by the protocol 
that a cache from the smallest enclosing domain containing a given data is always used. Locality is also exploited to reduce the latency in long distance
communication. With certain weak assumptions (see section \ref{subsec:latency}) about the topology of the underlying 
network, we can guarantee that distance travelled
by a message in HCA overlay network is linearly proportional to the optimal distance in the underlying network and $O(log(n))$ nodes
are visited along the path of the message. There are ways to optimize the constant factor of latency even more, but we presented the solution
in Subsection \ref{latopt}
because it would have decreased cache utilization in case of migrating writers of the shared data. 

HCA also supports a simple but effective model of access rights to distributed resources to protect resources owned by users. HCA should
be also secure from typical attacks against the network itself; the resources owned by users, like the persistent storage space, are secure 
even in the case of compromised communication nodes.
Reading rights of shared data and authentication of the source of information could be 
compromised by a hostile communication node. This is a real
possibility because the network is administered by multiple organizations. Encryption and authentication of data using
public key cryptography can be used on top of HCA to prevent these attacks but at the same time the access right management becomes
cumbersome and inefficient. Security could also be enhanced indirectly by logging usage history and thus increasing the traceability
and accountability of user actions.

\section{Critique}

HCA does not yet fully qualify as an ideal peer-to-peer architecture because the topology of the domain hierarchy is not formed
and improved automatically by the nodes. This compromise was reached mainly because service provider organizations were assumed to be
responsible for administering the communication nodes to achieve simplicity, efficiency, and security.
A good topology for the HCA domain hierarchy requires some pre-planning and a little initial
coordination between top organizations administering the system. Individual nodes also need some configuration information to know
where to attach to the network. The access control mechanism and integrity of the HCA also require a certain level of trust between 
communication nodes. A relatively good availability is expected of the nodes, which rules out the use of home-based nodes if 
good performance is required. 
Some sophisticated protocols have been developed to provide security in distributed peer-to-peer systems where
large portion of nodes can be hostile and, in theory, it could be possible to provide the functionality of HCA with completely unreliable
nodes that do not need basically any configuration.

Rich quality-of-service guarantees are not supported by HCA. On top of Internet any strict guarantees would be impossible to implement,
anyway, and combining scalability and fairness in the network resouce usage with QoS specifications requires further research.
The need for support for QoS guarantees is decreasing as the network bandwidth, latency, and reliability improve. In the end,
QoS will be needed only by special applications requiring absolute real-time guarantees of the delivery just like the applications
requiring a real-time operating system. 

HCA could be used for scalable delivery of web content, but this requires some kind of mapping between web addresses and HCA
socket references and implementation bridging, for example, a web proxy to HCA and tools for content providers to migrate their
web content to HCA. Because this application is outside the scope of this work we will just state the possibility of this
kind of direction for further work but will probably ignore it ourselves.

HCA does not provide any support for global consistency of the state of the shared data. SharedVectors states are updated in isolation
of other SharedVectors. Transactional semantics must, therefore, be implemented on top of HCA, but to implement the coordination at
a higher level is less efficient. The possibility to provide some global consistency would be a powerful feature if it can be combined
with scalability but this is a difficult problem and requires further research.

Anonymity of senders and receivers of information is not supported in the architecture of HCA in any way. With a proper choice of 
intra-domain protocol the HCA network cannot be shut down in whole from any single control point and a global censor could be difficult
to implement. However, it follows from the architecture of HCA that it is possible to censor any single known source and local domains
could be shut down entirely if locality is exploited in the topology. Therefore, the ideals of Freenet are difficult to 
integrate into the architecture in its current form and a lot of research would be needed to include these features.

Global monitoring of the activity of the HCA network is not supported in the protocol. If implementations try to solve this by themselves
we could lose the interoperability of different implementations of HCA. Also, any standard format for logging information, for example, about
usage statistics is not specified in the basic architecture. The lack of these features can be an important shortcoming in commercial
large scale use and also hamper testing and tuning of a global configuration. 

HCA does not support human readable names, meta-data, or trading services for finding and identification of sockets. 
This is purely a design choice motivated by modularity. Naming and finding of resources is an orthogonal aspect of the system
and, in this sense, should be implemented on platform level but, for example, the human readable names in itself are a difficult
problem, because names refer to external world and should be controlled by some central authority. Therefore we have separated
the location of a resource in HCA from the naming of the resource
which is left as a responsibility of an external naming service. Name-to-identifier and identifier-to-location mappings also have different 
requirements which can be taken into account better if the services are separated.
For example, in the case of real-time updates of the location of a resource, logical indentifier should always point to the current location of
the resource but names could be updated more slowly.

In the first version of HCA we haven't specified any operations for changing the domain tree structure while the system is running.
It is possible to add and remove nodes inside a domain but, for example, moving a whole subtree under a new parent domain is
not possible. This is a serious weakness because the changes in underlying network and administering organizations are likely to
cause a need for a change in the HCA topology without interfering with the normal use of the system. 

\section{Comparison to Other Technologies}

A rough overview of the most important features that differentiate HCA and other
technologies introduced in Chapter \ref{ch:problemDef} are collected into Table \ref{tab:techComp}. 
The systems all have somewhat different intended use but nevertheless share many techniques.

\begin{sidewaystable}
\centering
\caption{Technology comparison}
\scriptsize
\begin{tabular}{|p{2cm}|p{3cm}p{3cm}p{2,5cm}p{4cm}p{4cm}|}
\hline
Feature & BitTorrent & Chord & IP multicast & Globe location service & HCA \\
\hline
intended use & file sharing & distributed lookup of a node associated with a given key & N-to-N unreliable communication of messages & location of migrating objects & N-to-1 communication of messages and 1-to-N data sharing \\
\hline
locality & no & improved in later versions & yes & yes & yes\\
\hline
streaming & no & no & yes & no & yes \\
\hline
state & stateful & stateful & stateless & stateful & stateful\\
\hline
scalability & almost & yes & yes & yes & yes \\
\hline
fault-tolerance & almost & excellent & yes & yes & yes \\
\hline
distribution transparencies & failure, replication & location, failure & location & location, migration, failure, replication & location, migration, failure, replication\\
\hline
security & no & no & no & assumes single operating organization, basic access control except for lookup & basic access control, should tolerate many type of attacks \\
\hline
reliability & yes & yes & no & yes & yes \\
\hline
maintenance & file provider must provide tracker and .torrent file & easy, completely decentralized architecture & managed by service providers
& managed by a support organization, supports dynamically changing the search tree topology & managed by a group of support organizations working together, preplanning of topology and some manual per node configuration required\\
\hline
complexity & simple & simple & fairly simple & complex & fairly complex\\
\hline
\end{tabular}
\normalsize
\label{tab:techComp}
\end{sidewaystable}

The first row in the Table gives a short description of the intended use of each technology. On the second row
the locality feature refers to the ability of the system to use location information in optimizing latency in the system.
Streaming feature means possibility to use push semantics in data sharing and support for small cache updates. State
refers to the communication system inner state: this can be either stateless or stateful. Scalability can be
interpreted here to mean the ability of the system to support growing number of users with same
architecture and reasonable cost. Fault-tolerance refers to the systems ability to recover from errors. Distribution
transparencies row lists for each technology which aspects of distribution the system hides from its clients. Security
feature means how robust the system is against malicious use, that is, can the original function of the system sustained
under attacks and if the implementation remains usable to legitimate clients. Reliability row lists whether information can be lost or changed
in communication without notifation to the receiver. For example, IPv6 implements unreliable multi-cast, which means that it doesn't
guarantee that every receiver gets all messages. Maintenance row contains some kind of estimate of the work needed to setup the system and
keep it working in its intended use. Requirements for the environment of building each system are also listed on the maintenance row.
Complexity means here a rough qualitative estimate by the author of the 
complexity of each system's architecture, interfaces, algorithms, and protocols combined.

BitTorrent is considered almost scalable, because its architecture has the central tracker
component that does not scale but that is not a practical problem with normal use of the system because the overhead communication
with the tracker is only one thousandth part of the bandwidth used by peers. Also, the random graphs connecting peers in the BitTorrent
are very robust against failures between peers but the tracker is also a single point of failure in this respect.

From the comparison chart it is easy to see that the intended use of the system dictates which features are important.
We could not find any system with exactly same goals as HCA, but the basic features of HCA can be individually 
found in many other systems. There are some technical details of HCA, which we have not found in other
systems like the protocol of shortcut connections but the real contribution of the HCA is the integration of a new combination of known 
techniques like DHTs, domain hierarchies for achieving locality, and multi-cast based data sharing and streaming with access rights. 
We believe that the combination of features in HCA enables the building of a high-level distribution middleware in an
efficient and architecturally sound way.
   \chapter{CONCLUSION}
\label{ch:conclusion}

In this thesis, we presented a software architecture with a novel combination of features, algorithms, 
and a protocol for building a general-purpose scalable
and fault-tolerant communication platform. The system provides natively both sharing and streaming of data and 
message-based communication in a location-transparent way. The main goals were attained and the results
were evaluated. There is still a lot of 
development work to do and the architecture needs to be fully implemented and tested in an actual distributed setting
before a final evaluation can be done.

The scalable sharing of data and distribution middlewares are hot research topics at the moment and 
many similar projects are being developed simultaneously. This interest is motivated by numerous potential applications of 
the Internet and the vision of a possible
shift in how software is developed to the networked computers. Therefore,
the topic is considered to be of great significance: a combination of a robust implementation and state-of-the-art 
results of theoretical research could have a noticeable impact.

\subsubsection{Future Plans}

The author will continue the development of HCA by implementing a complete prototype
of the design. After finishing the prototype implementation a comprehensive testing in a global distributed
setting will be carried out. The design can then be iterated and the implementation tuned based on the test results.
Hopefully, the author will get funding or outside help so that the testing can be done on a large
enough scale. The prototype will then be developed to the level of robustness and efficiency needed
in commercial applications. It is also possible that features, such as logging statistics of network use and tools for
socket file management, are needed in this phase. 
After all, right from the beginning the goal has been to produce
a realistic system for real-life applications instead of just research.

The questions about quality of service specifications, domain tree management and ease of setup will be further researched. There are lots
of other possible directions for further development like logging and collecting statistics about network
usage, integration to web, naming and location of files, and anonymity of users, but these research directions 
are not essential at this time.

As already suggested in Chapter \ref{introduction}, HCA is designed to be only the first layer of a larger
distributed platform, although it has value as a separate component, too. The next step in this bigger project
is a development of a layer that raises the abstraction level of HCA. It will be a type of distributed operating
system based on a high-level functional programming language that uses HCA for distribution.




\bibliographystyle{plain}

\renewcommand{\bibname}{List of References}
\addcontentsline{toc}{chapter}{\quad\,\,{List of References}}
\bibliography{References}

\begin{thebibliography}{10}

\bibitem{GLOBELOC}
Gerco Ballintijn.
\newblock {\em Locating Objects in a Wide-area System}.
\newblock PhD thesis, Vrije Universiteit Amsterdam, 2003.

\bibitem{GLOBELOCSCALE}
Gerco Ballintijn and Maarten van Steen.
\newblock Exploiting location awareness for scalable location-independent
  object {ID}s.
\newblock In {\em Fifth Annual ASCI Conference}, pages 321--328, 1999.

\bibitem{HERODOTUS}
Timo Burkard.
\newblock Herodotus: A peer-to-peer web archival system.
\newblock Master's thesis, Massachusetts Institute of Technology, 2002.

\bibitem{FREENET}
Ian Clarke, Oskar Sandberg, Brandon Wiley, and Theodore~W. Hong.
\newblock {Freenet}: A distributed anonymous information storage and retieval
  system.
\newblock In {\em Workshop on Design Issues in Anonymity and Unobservability},
  pages 46--66, July 2000.

\bibitem{BITTORRENT}
Bram Cohen.
\newblock Incentives build robustness in {BitTorrent}.
\newblock Available from http://www.bittorrent.com/bittorrentecon.pdf, 2003.

\bibitem{MTP}
Jon Crowcroft and Karen Paliwoda.
\newblock A multicast transport protocol.
\newblock In {\em SIGCOMM '88: Symposium proceedings on Communications
  architectures and protocols}, pages 247--256, New York, NY, USA, 1988. ACM
  Press.

\bibitem{CFS}
Frank Dabek.
\newblock A cooperative file system.
\newblock Master's thesis, Massachusetts Institute of Technology, 2001.

\bibitem{IPv6}
Stephen Deering and Robert Hinden.
\newblock Internet protocol, version 6 (ipv6), 1998.
\newblock IETF, RFC 2460.

\bibitem{DCOM}
Guy Eddon and Henry Eddon.
\newblock {\em Inside Distributed {COM}}.
\newblock Microsoft Press, 1998.

\bibitem{PATTERNS}
Erich Gamma, Richard Helm, Ralph Johnson, and John Vlissides.
\newblock {\em Design Patterns: Elements of Reusable Object-Oriented Software}.
\newblock Addison-Wesley Publishing Company, Inc., 1995.

\bibitem{TRANSACTIONS}
Jim Gray and Andreas Reuter.
\newblock {\em Transaction Processing: Concepts and Techniques}.
\newblock Morgan Kaufmann, 1993.

\bibitem{hagino87}
Tatsuya Hagino.
\newblock {\em A Categorical Programming Language}.
\newblock PhD thesis, University of Edinburgh, 1987.

\bibitem{BITTORRENTCHEAT}
David Hales and Simon Patarin.
\newblock How to cheat {BitTorrent} and why nobody does.
\newblock Technical Report UBLCS-2005-12, University of Bologna, May 2005.
\newblock Available from http://www.cs.unibo.it/pub/TR/UBLCS/2005/2005-12.pdf.

\bibitem{Globe}
Philip Homburg.
\newblock {\em The Architecture of a Worldwide Distributed System}.
\newblock PhD thesis, Vrije Universiteit Amsterdam, 2001.
\newblock Also available from http://www.cs.vu.nl/globe/.

\bibitem{OSI}
{International Organization for Standardization (ISO)}.
\newblock {Open systems Interconnection (OSI)}.

\bibitem{HASKELL}
Simon~Peyton Jones.
\newblock {\em Haskell 98 Language and Libraries}.
\newblock Cambridge University Press, 2003.

\bibitem{DYNHUFF}
Donald Knuth.
\newblock Dynamic huffman coding.
\newblock {\em Journal of Algorithms}, 6:163--180, 1985.

\bibitem{OLIOKIRJA}
Kai Koskimies.
\newblock {\em Oliokirja}.
\newblock Satku, 1998.

\bibitem{DOTNET}
{Microsoft Corporation}.
\newblock {Microsoft .NET}.
\newblock homepage at http://www.microsoft.com/net/.

\bibitem{ML}
Robin Milner, Mads Tofte, Robert Harper, and David MacQueen.
\newblock {\em The Definition of Standard {ML} - Revised}.
\newblock The MIT Press, 1997.

\bibitem{CORBA}
Common object request broker architecture, version 3.0.3.
\newblock Core Specification, March 2004.
\newblock Available 16.5.2006 from http://www.omg.org.

\bibitem{POPULARITY}
Andrew Parker.
\newblock The true picture of peer-to-peer filesharing, 2004.
\newblock Presentation available 16.5.2006 from
  http://www.cachelogic.com/research/slide1.php.

\bibitem{poll98subtyping}
Erik Poll.
\newblock Subtyping and inheritance for categorical datatypes (preliminary
  report).
\newblock In {\em Proceedings Meeting on Theories for Types and Proofs,
  {TTP}-Kyoto'97}, volume 1023. Kyoto Univ.\ Research Inst.\ for Math.\
  Sciences, Kyoto, 1998.

\bibitem{UDP}
User datagram protocol, 1980.
\newblock IETF, RFC 768.

\bibitem{IP}
Internet protocol, 1981.
\newblock IETF, RFC 791.

\bibitem{TCP}
Transmission control protocol, 1981.
\newblock IETF, RFC 793.

\bibitem{rei95}
Horst Reichel.
\newblock An approach to object semantics based on terminal coalgebras.
\newblock {\em Mathematical Structures in Computer Science}, 5:129--152, 1995.

\bibitem{ERASURE}
Rodrigo Rodrigues and Barbara Liskov.
\newblock High availability in {DHT}s: Erasure coding vs. replication, 2005.
\newblock Available from
  http://iptps05.cs.cornell.edu/PDFs/CameraReady{\_}223.pdf.

\bibitem{MBONE}
Kevin Savetz, Neil Randall, and Yves Lepage.
\newblock {\em {MBONE}: Multicasting Tomorrow's Internet}.
\newblock John Wiley \& Sons Inc (Computers), 1996.

\bibitem{CHORD}
Ion Stoica, Robert Morris, David Karger, M.~Frans Kaashoek, and Hari
  Balakrishnan.
\newblock {Chord}: A scalable peer-to-peer lookup service for internet
  applications.
\newblock In {\em ACM SIGCOMM 2001}, pages 149--160, 2001.
\newblock Available from http://pdos.csail.mit.edu/chord/\#pubs.

\bibitem{JAVA}
{Sun Microsystems, Inc.}
\newblock {Java 2 platform Enterprise Edition}.
\newblock homepage at http://java.sun.com/j2ee.

\bibitem{steen01achieving}
Maarten van Steen and Gerco Ballintijn.
\newblock Achieving scalability in hierarchical location services, 2001.
\newblock Maarten van Steen and Gerco Ballintijn. Achieving Scalability in
  Hierarchical Location Services. Technical Report {IR-491}, Vrije
  Universiteit, Department of Mathematics and Computer Science, Nov. 2001.

\end{thebibliography}

     \appendix



\end{document}